\documentclass[prd,aps,preprint,showpacs,nofootinbib,superscriptaddress]{revtex4-1}
\usepackage{epsfig}
\usepackage{amsmath}
\usepackage{color}
\usepackage{latexsym}
\usepackage{amssymb}
\usepackage{dsfont}
\usepackage{multirow}

\newcommand{\bea}{\begin{eqnarray}}
\newcommand{\eea}{\end{eqnarray}}
\newcommand{\be}{\begin{equation}}
\newcommand{\ee}{\end{equation}}

\newcommand{\ud}{\mathrm{d}}
\newcommand{\uL}{\mathcal{L}}

\newcommand{\uTr}{\mathrm{Tr}}

\newcommand{\barpsi}{\overline{\psi}}

\newlength\savedwidth

\begin{document}

\title{Geometrical approach to the proton spin decomposition}

\author{C\'edric Lorc\'e}
\affiliation{IPNO, Universit\'e Paris-Sud, CNRS/IN2P3, 91406 Orsay, France\\
 and LPT, Universit\'e Paris-Sud, CNRS, 91406 Orsay, France\\
\emph{e-mail}: lorce@ipno.in2p3.fr}

\begin{abstract}
We discuss in detail and from the geometrical point of view the issues of gauge invariance and Lorentz covariance raised by the approach proposed recently by Chen \emph{et al.} to the proton spin decomposition. We show that the gauge invariance of this approach follows from a mechanism similar to the one used in the famous Stueckelberg trick. Stressing the fact that the Lorentz symmetry does not force the gauge potential to transform as a Lorentz four-vector, we show that the Chen \emph{et al.} approach is Lorentz covariant provided that one uses the suitable Lorentz transformation law. We also make an attempt to summarize the present situation concerning the proton spin decomposition. We argue that the ongoing debates concern essentially the physical interpretation and are because of the plurality of the adopted pictures. We discuss these different pictures and propose a pragmatic point of view.
\end{abstract}

\maketitle

\section{Introduction}

It was a big surprise when the experimental results of the European Muon Collaboration (EMC) showed that only a small fraction of the proton spin is carried by the quark spin \cite{Ashman:1987hv,Ashman:1989ig}, in clear contradiction with the naive quark model picture where the proton spin originates solely from the quark spins. Even a less naive picture, where the three constituent quarks are allowed to orbit, cannot explain such a small fraction. This triggered the so-called proton ``spin puzzle'' which is one of the most intriguing and interesting topics of hadronic physics \cite{Leader:1988vd}. A lot of effort has then been made on both theoretical and experimental sides to define and access the missing pieces of the puzzle. According to recent analyses, it turns out that about $1/3$ of the nucleon spin comes from the quark spin \cite{Ageev:2005gh,Alexakhin:2006vx,Airapetian:2007mh} while the gluon spin seems to contribute little \cite{Ageev:2005pq,Boyle:2006ab,Kiryluk:2005mh,Fatemi:2006aa}, see also the short reviews \cite{Kuhn:2008sy,Burkardt:2008jw}. These results increased in particular the interest in the orbital angular momentum (OAM) which should account for the substantial missing contribution.

A decade ago, there were essentially two popular decompositions of the proton spin: one is the Jaffe-Manohar decomposition \cite{Jaffe:1989jz} and the other is the Ji decomposition \cite{Ji:1996ek}. The former has a simple partonic interpretation and provides a complete decomposition into quark spin, quark OAM, gluon spin and gluon OAM contributions. However, it is not gauge invariant and is then considered in the light-front gauge in order to make contact with the parton model picture. Later, Bashinsky and Jaffe \cite{Bashinsky:1998if} proposed a variation of the Jaffe-Manohar decomposition which has the virtue of being invariant under the residual gauge symmetry. On the contrary, the Ji decomposition is gauge invariant. However, it has no simple partonic interpretation and does not provide any decomposition of the gluon total angular momentum into spin and OAM contributions, in agreement with the textbook claim that there exists no local gauge-invariant operator for the gluon spin \cite{Jauch, Berestetskii}. 

Recently, Chen \emph{et al.} \cite{Chen:2008ag,Chen:2009mr} proposed to separate explicitly the gauge potential into pure-gauge and physical parts. They obtained a gauge-invariant decomposition which reduces to the Jaffe-Manohar decomposition in a non-abelian generalization of the Coulomb gauge. This new approach triggered many theoretical works in the last few years \cite{Tiwari:2008nz,Ji:2010zza,Ji:2009fu,Ji:2012gc,Chen:2008gv,Chen:2008ja,Chen:2009dg,Chen:2011zzh, Wong:2010rs,Wang:2010ao,Chen:2011gn,Chen:2012vg,Goldman:2011vs,Stoilov:2010pv,Wakamatsu:2010qj,Wakamatsu:2010cb,Wakamatsu:2011mb,Wakamatsu:2012ve,Cho:2010cw,Cho:2011ee,Hatta:2011zs,Hatta:2011ku,Hatta:2012cs,Zhang:2011rn,Leader:2011za,Lorce:2012ce}, and raised a lot of criticism, especially regarding the questions of gauge invariance and Lorentz covariance, see \emph{e.g.} \cite{Tiwari:2008nz,Ji:2010zza,Ji:2009fu}. Chen \emph{et al.} basically replied to this criticism \cite{Chen:2008gv,Chen:2008ja,Chen:2009dg}, but were not sufficiently convincing, as one can see from \emph{e.g.} Ref. \cite{Ji:2012gc}. For this reason, we come back to these questions in greater details.

This paper is divided into two parts:
\begin{itemize}
\item The first part is a bit technical but shows explicitly that the Chen \emph{et al.} approach is gauge invariant and is consistent with the Lorentz covariance. In section \ref{GaugeSection} we remind the basics of gauge symmetry emphasizing its geometrical interpretation. We provide a geometrical interpretation of the Chen \emph{et al.} approach, and show that it is based on a mechanism similar to the one used in the famous Stueckelberg trick. In section \ref{LorentzSection}, we discuss the Lorentz transformation laws in a gauge theory and show that, contrary to a widespread belief, the gauge potential does not necessarily transform as a Lorentz four-vector. Then we conclude that the Chen \emph{et al.} approach is Lorentz covariant provided that one works with the suitable representation of the Poincar\'{e} group. 
\item The second part of this paper starts with section \ref{DecompositionSection}, where we summarize and compare the main decompositions of the proton spin. We discuss in section \ref{DiscussionSection} the different points of view and show how they affect the ongoing controversies. We recommend the adoption of a pragmatic point of view, and discuss how one can in principle access the different kinds of OAM.
\end{itemize}
 Finally, we conclude this paper with section \ref{ConclusionSection}.

\section{Gauge invariance}\label{GaugeSection}

In this section, we remind the geometrical picture behind gauge theories, though in a slightly non-standard way, which stresses the similarities with general relativity\footnote{In the language of differential geometry, it corresponds to the similarities between the fiber bundles and the tangent bundle.}, see \emph{e.g.} Refs. \cite{Frankel,Ryder,Peskin,Moriyasu,Utiyama:1956sy}. We then discuss the approach proposed by Chen \emph{et al.} and provide its geometrical interpretation. Finally, we show that the gauge invariance of the Chen \emph{et al.} approach follows from a mechanism similar to the one used in the famous Stueckelberg trick.

\subsection{Geometrical picture}

In gauge theories, a copy $\mathcal V_x$ of the internal space is attached to each space-time point $x$. The source field $\psi$ then specifies a vector in each copy of the internal space. A gauge transformation corresponds to a change of basis in the internal space. Under such a transformation, the components of the source field naturally transform as those of an internal vector
\begin{equation}\label{gaugetf}
\psi(x)\mapsto\tilde\psi(x)=U(x)\psi(x),
\end{equation}
where $U(x)$ is a unitary matrix in the internal space\footnote{One usually considers only changes that conserve the orthonormality of the basis.}. The $x$-dependence of this transformation law indicates that the change of basis can be different in each copy of the internal space. The gauge symmetry principle states that physics does not depend on the particular choice of basis in each $\mathcal V_x$, \emph{i.e.} the physical quantities have to be invariant under gauge transformations. The gauge symmetry is therefore not a physical symmetry in the sense that it is \emph{exact} and therefore not observable. It can then be considered as a mere redundancy of the mathematical description of the physical system.

One usually needs to compare the source fields at two infinitely close (but separate) space-time points $x$ and $x+\ud x$, \emph{i.e.} vectors belonging to different copies of the internal space. One therefore needs to introduce a (parallel transport) rule which maps vectors in $\mathcal V_{x+\ud x}$ onto vectors in $\mathcal V_x$ 
\begin{equation}
\psi(x+\ud x)\mapsto\psi_\parallel(x+\ud x)=\left[\mathds 1-igA_\mu(x)\,\ud x^\mu\right]\psi(x+\ud x),
\end{equation}
where the so-called gauge potential field $A_\mu(x)$ is a connection defining the notion of parallelism, and is the analogue of the Christoffel symbols in general relativity. The intrinsic variation of the source field is then given by
\begin{align}
\psi_\parallel(x+\ud x)-\psi(x)&=\psi(x+\ud x)-\psi(x)-igA_\mu(x)\psi(x)\,\ud x^\mu\nonumber\\
&=\left[\partial_\mu-igA_\mu(x)\right]\psi(x)\,\ud x^\mu\nonumber\\
&=D_\mu\psi(x)\,\ud x^\mu,
\end{align}
where $D_\mu\equiv\partial_\mu-igA_\mu$ is called the covariant derivative. By construction, the covariant derivative of an internal vector is a vector belonging to the same copy of the internal space. In other words, the covariant derivative of a source field transforms like a source field
\begin{equation}
D_\mu\psi(x)\mapsto\widetilde{D_\mu\psi}(x)=U(x)D_\mu\psi(x).
\end{equation}
 The gauge transformation laws of the covariant derivative and of the connection are then given by
\begin{align}
D_\mu&\mapsto\tilde D_\mu=U(x)D_\mu U^{-1}(x),\label{covgauge}\\
A_\mu(x)&\mapsto\tilde A_\mu(x)=U(x)\left[A_\mu(x)+\frac{i}{g}\,\partial_\mu\right]U^{-1}(x).\label{nonabelian}
\end{align}
In particular, for electrodynamics one has $g=-e$ and $U(x)=e^{-ie\alpha (x)}$ with $\alpha$ an arbitrary function of space and time, so that the gauge transformation law \eqref{nonabelian} reduces to the familiar abelian one
\begin{equation}\label{abelian}
A^\mu(x)\mapsto \tilde A^\mu(x)=A^\mu(x)+\partial^\mu\alpha(x).
\end{equation}
Note also that, contrary to the covariant derivative, the gauge potential does not transform as an internal tensor because of the extra term $\frac{i}{g}\,U\partial_\mu U^{-1}$, typical of a connection.

Contrary to ordinary derivatives, the covariant derivatives do not commute with each other. Their commutator defines the so-called field strength tensor
\begin{equation}
F_{\mu\nu}\equiv\frac{i}{g}[D_\mu,D_\nu]=\partial_\mu A_\nu-\partial_\nu A_\mu-ig[A_\mu,A_\nu],
\end{equation}
which transforms as an internal tensor
\begin{equation}
F_{\mu\nu}(x)\mapsto\tilde F_{\mu\nu}(x)=U(x)F_{\mu\nu}(x)U^{-1}(x),
\end{equation}
as one can see directly from eq.~\eqref{covgauge}.  It is the analogue of the Riemann curvature tensor in general relativity and can be thought of as a tensor describing some sort of internal curvature.

\subsection{Physical and pure-gauge degrees of freedom}

The gauge potential has four components in the physical space, but it doesn't mean that it has four physical degrees of freedom. In fact, the time component is not dynamical since the Lagrangian does not contain its time derivative, and one degree of freedom is decoupled because of the invariance of the theory under gauge transformations. The gauge potential has therefore only two physical degrees of freedom corresponding to the two physical polarizations $h=\pm 1$ of the associated gauge boson, namely the photon in QED and the gluon in QCD. 

The redundancy in the mathematical description implied by the gauge symmetry is source of many theoretical difficulties. Tto avoid these complications, one can remove the gauge freedom from the very beginning by imposing conditions on the gauge potential. A typical example is the quantization of electrodynamics in the Coulomb gauge, see \emph{e.g.} \cite{BjorkenDrell,Manoukian:1987hy}. The advantage of this approach is that one is working only with the physical degrees of freedom, but the price to pay is that one looses \emph{explicit} Lorentz covariance and gauge invariance \cite{Strocchi:1974xh}. 

Chen \emph{et al.} proposed a different approach \cite{Chen:2008ag,Chen:2009mr}. The idea is to separate explicitly the unphysical \emph{pure-gauge} part of the gauge potential from the \emph{physical} part. By definition, the pure-gauge part of the potential has the same gauge transformation law as the full gauge potential
\begin{equation}\label{Apure}
A_\mu^\text{pure}(x)\mapsto\tilde A_\mu^\text{pure}(x)=U(x)\left[A_\mu^\text{pure}(x)+\frac{i}{g}\,\partial_\mu\right]U^{-1}(x),
\end{equation}
and does not contribute to the field strength 
\begin{equation}
F_{\mu\nu}^\text{pure}\equiv\partial_\mu A^\text{pure}_\nu-\partial_\nu A^\text{pure}_\mu-ig[A^\text{pure}_\mu,A^\text{pure}_\nu]=0.
\end{equation}
The physical part is then defined as the complement\footnote{Note that to make this approach concrete, one has to impose further constraints on $A^\text{phys}_\mu$ because of the Stueckelberg symmetry discussed in the next section.}
\begin{equation}
A^\text{phys}_\mu\equiv A_\mu-A_\mu^\text{pure}.
\end{equation} 
Note in particular that, even though $F_{\mu\nu}^\text{pure}=0$, one has in non-abelian gauge theories
\begin{equation}
F_{\mu\nu}\neq\partial_\mu A^\text{phys}_\nu-\partial_\nu A^\text{phys}_\mu-ig[A^\text{phys}_\mu,A^\text{phys}_\nu].
\end{equation}
It follows from the definition of $A_\mu^\text{pure}$ that the physical part of the gauge potential transforms as an internal tensor
\begin{equation}\label{Aphys}
A_\mu^\text{phys}(x)\mapsto\tilde A_\mu^\text{phys}(x)=U(x)A_\mu^\text{phys}(x)U^{-1}(x).
\end{equation}
In practice, the qualifier\footnote{The qualifier ``physical'' is unfortunate as it seems to imply also uniqueness. It became however standard in the literature, so I decided to stick to it.} ``physical'' is synonymous with ``tensor under (ordinary) gauge transformations''. The field strength tensor and the source field are then other examples of physical fields. One would also like to require that $A_\mu^\text{phys}$ contains only the physical degrees of freedom. Since a \emph{physical} gauge condition removes all gauge freedom, there should exist a gauge transformation such that $\tilde A_\mu=\tilde A^\text{phys}_\mu$ and therefore $\tilde A_\mu^\text{pure}=0$. Consequently, one can write in general $A_\mu^\text{pure}$ as a pure-gauge term
\begin{equation}
A_\mu^\text{pure}=\frac{i}{g}\,U_\text{pure}\partial_\mu U^{-1}_\text{pure}.
\end{equation}
Obviously, such a term cannot contribute to the field strength.

From a geometrical point of view, the Chen \emph{et al.} approach amounts to assume that there exists some privileged or ``natural'' basis in each copy of the internal space. In the following, we will denote the components of any internal tensor in this natural basis with a hat. One can then write the source field as
\begin{equation}
\psi=U_\text{pure}\hat\psi,
\end{equation}
where $U_\text{pure}$ is the internal rotation which relates the components $\hat\psi$ of the source field in the natural basis to the components $\psi$ in an arbitrary basis. Manifestly, only $U_\text{pure}$ is affected by a gauge transformation
\begin{align}
\hat\psi(x)&\mapsto\tilde{\hat\psi}(x)=\hat\psi(x),\\
U_\text{pure}(x)&\mapsto\tilde U_\text{pure}(x)=U(x)U_\text{pure}(x).\label{leftU}
\end{align}
The natural variation of the source field corresponds to the variation of its components in the natural basis $\hat\psi(x+\ud x)-\hat\psi(x)$. Expressed in an arbitrary internal basis, it reads
\begin{align}
U_\text{pure}(x)\left[\hat\psi(x+\ud x)-\hat\psi(x)\right]&=U_\text{pure}(x)\partial_\mu\hat\psi(x)\,\ud x^\mu\nonumber\\
&=U_\text{pure}(x)\partial_\mu\left[U^{-1}_\text{pure}(x)\psi(x)\right]\ud x^\mu\nonumber\\
&=D_\mu^\text{pure}\psi(x)\,\ud x^\mu,
\end{align}
where $D^\text{pure}_\mu\equiv\partial_\mu-igA^\text{pure}_\mu$ is called the pure-gauge covariant derivative. The assumption of a natural basis provides us directly with a natural connection $A^\text{pure}_\mu$. Note that, contrary to the ordinary covariant derivatives, the pure-gauge covariant derivatives commute with each other $[D^\text{pure}_\mu,D^\text{pure}_\nu]=-igF^\text{pure}_{\mu\nu}=0$. So, in this approach, the internal space is not considered as curved.

Both the strength field tensor and the physical part of the gauge potential are internal tensors, and therefore have also natural components
\begin{align}
F_{\mu\nu}&=U_\text{pure}\hat F_{\mu\nu}U^{-1}_\text{pure},\\
A_\mu^\text{phys}&=U_\text{pure}\hat A_\mu^\text{phys}U^{-1}_\text{pure}.
\end{align}
Note in particular that one has the welcome feature
\begin{equation}
\hat F_{\mu\nu}=\partial_\mu \hat A^\text{phys}_\nu-\partial_\nu \hat A^\text{phys}_\mu-ig[\hat A^\text{phys}_\mu,\hat A^\text{phys}_\nu].
\end{equation}
Similarly to the source field, one can consider the natural variation of the physical part of the gauge potential $\hat A^\text{phys}_\nu(x+\ud x)-\hat A^\text{phys}_\nu(x)$. Expressed in an arbitrary internal basis, it reads
\begin{align}
U_\text{pure}(x)&\left[\hat A^\text{phys}_\nu(x+\ud x)-\hat A^\text{phys}_\nu(x)\right]U^{-1}_\text{pure}(x)\nonumber\\
&=U_\text{pure}(x)\partial_\mu\hat A^\text{phys}_\nu(x)U^{-1}_\text{pure}(x)\,\ud x^\mu\nonumber\\
&=U_\text{pure}(x)\partial_\mu\left[U^{-1}_\text{pure}(x)A_\nu^\text{phys}(x)U_\text{pure}(x)\right]U^{-1}_\text{pure}(x)\,\ud x^\mu\nonumber\\
&=\mathcal D_\mu^\text{pure}A_\nu^\text{phys}(x)\,\ud x^\mu,
\end{align}
where $\mathcal D^\text{pure}_\mu\equiv\partial_\mu-ig\left[A^\text{pure}_\mu,\quad\right]$ is called the \emph{adjoint representation} pure-gauge covariant derivative. Similar covariant derivatives can be obtained for any other representation, just like in general relativity. Thanks to this new pure-gauge covariant derivative, one can simply relate the field strength tensor to the physical part of the gauge potential
\begin{equation}
F_{\mu\nu}=\mathcal D^\text{pure}_\mu A^\text{phys}_\nu-\mathcal D^\text{pure}_\nu A^\text{phys}_\mu-ig[A^\text{phys}_\mu,A^\text{phys}_\nu].
\end{equation}

\subsection{Stueckelberg gauge symmetry}\label{Stuecksect}

It is well-known that the introduction of a mass term for the photon breaks explicitly the $U(1)$ gauge symmetry. However, Stueckelberg found a mechanism where such a term can be introduced without breaking the gauge invariance \cite{Stueckelberg:1900zz,Stueckelberg:1938zz,Stueckelberg:1957zz}. The idea consists in increasing the number of fields without increasing the number of degrees of freedom thanks to an additional symmetry \cite{Pauli:1941zz,Ruegg:2003ps}. As observed by Stoilov \cite{Stoilov:2010pv}, the Chen \emph{et al.} approach is based on a similar mechanism: the separation of the gauge potential into pure-gauge and physical parts (\emph{i.e.} instead of one field $A_\mu$, one plays with two fields $A_\mu^\text{pure}$ and $A_\mu^\text{phys}$) leads to an enlarged gauge symmetry. In the case of QED, on top of the electromagnetic $U(1)_{EM}$ gauge symmetry \eqref{abelian}, there is an additional $U(1)_S$ gauge symmetry referred to as the \emph{Stueckelberg} symmetry\footnote{Writing the (abelian) pure-gauge field as $A_\mu^\text{pure}(x)=\partial_\mu\alpha^\text{pure}(x)$, one sees that the scalar function $\alpha^\text{pure}(x)$ plays a role similar to the Stueckelberg field $B(x)/m$. Note that contrary to the Stueckelberg mechanism, the function $C(x)$ does not need to satisfy the massive Klein-Gordon equation.} 
\begin{align}
A_\mu^\text{pure}(x)&\mapsto A_\mu^{\text{pure},g}(x)=A_\mu^\text{pure}(x)-\partial_\mu C(x),\label{stueckab1}\\
A_\mu^\text{phys}(x)&\mapsto A_\mu^{\text{phys},g}(x)=A_\mu^\text{phys}(x)+\partial_\mu C(x),\label{stueckab2}
\end{align}
where $C$ is an arbitrary scalar function of space and time. The full gauge group is therefore the direct product $U(1)_{EM}\times U(1)_{S}$. The Stueckelberg symmetry implies in particular that the pure-gauge condition $F_{\mu\nu}^\text{pure}=0$ is not sufficient to determine uniquely the decomposition $A_\mu=A_\mu^\text{pure}+A_\mu^\text{phys}$. 

From a geometrical point of view, the Stueckelberg symmetry corresponds to a change of natural basis without changing the actual basis used in the internal space
\begin{align}
\psi(x)&\mapsto\psi^g(x)=\psi(x),\\
\hat\psi(x)&\mapsto\hat\psi^g(x)=U_0(x)\hat\psi(x),\\
U_\text{pure}(x)&\mapsto U^g_\text{pure}(x)=U_\text{pure}(x)U^{-1}_0(x).\label{rightU}
\end{align}
Consequently, the Stueckelberg symmetry group is a copy of the original gauge group, and the full gauge group is simply the direct product of these two groups. Note that the original gauge transformation acts on the left of $U_\text{pure}$, see Eq. \eqref{leftU}, while the Stueckelberg symmetry acts on the right of it, see Eq. \eqref{rightU}. Obviously, the Stueckelberg symmetry does not affect the gauge potential $A_\mu$ but only its decomposition into pure-gauge and physical parts
\begin{align}
A_\mu(x)&\mapsto A^g_\mu(x)=A_\mu(x)\label{stueck}\\
A^\text{pure}_\mu(x)&\mapsto A^{\text{pure},g}_\mu(x)=A^\text{pure}_\mu(x)+\frac{i}{g}\,U_\text{pure}(x)U_0^{-1}(x)\left[\partial_\mu U_0(x)\right]U^{-1}_\text{pure}(x),\label{stueck1}\\
A^\text{phys}_\mu(x)&\mapsto A^{\text{phys},g}_\mu(x)=A^\text{phys}_\mu(x)-\frac{i}{g}\,U_\text{pure}(x)U_0^{-1}(x)\left[\partial_\mu U_0(x)\right]U^{-1}_\text{pure}(x).\label{stueck2}
\end{align}
For electrodynamics one has $g=-e$ and $U_0(x)=e^{-ieC(x)}$, so that the Stueckelberg transformation laws \eqref{stueck1} and \eqref{stueck2} reduce to the abelian ones \eqref{stueckab1} and \eqref{stueckab2}, respectively. In terms of natural components, the Stueckelberg transformation laws \eqref{stueck}-\eqref{stueck2} read
\begin{align}
\hat A_\mu(x)&\mapsto \hat A^g_\mu(x)=U_0(x)\left[\hat A_\mu(x)+\frac{i}{g}\,\partial_\mu\right]U^{-1}_0(x),\\
\hat A^\text{pure}_\mu(x)&\mapsto \hat A^{\text{pure},g}_\mu(x)=U_0(x)\hat A^\text{pure}_\mu(x)U^{-1}_0(x),\label{purestueck}\\
\hat A^\text{phys}_\mu(x)&\mapsto \hat A^{\text{phys},g}_\mu(x)=U_0(x)\left[\hat A^\text{phys}_\mu(x)+\frac{i}{g}\,\partial_\mu\right]U^{-1}_0(x).
\end{align}
Clearly, the field strength tensor $F_{\mu\nu}$  is invariant under Stueckelberg transformations, contrary to its expression in natural components
\begin{equation}
\hat F_{\mu\nu}(x)\mapsto \hat F^g_{\mu\nu}(x)=U_0(x)\hat F_{\mu\nu}(x)U^{-1}_0(x).
\end{equation}
Note also that one has 
\begin{equation}
F^{\text{pure},g}_{\mu\nu}\equiv\partial_\mu A^{\text{pure},g}_\nu-\partial_\nu A^{\text{pure},g}_\mu-ig\left[A^{\text{pure},g}_\mu,A^{\text{pure},g}_\nu\right]=0,
\end{equation}
showing that a pure-gauge term remains a pure gauge under Stueckelberg transformations. Consequently, Stueckelberg transformations map physical fields to physical fields. Moreover, since in the natural basis $\hat A^\text{pure}_\mu=0$, Eq. \eqref{purestueck} ensures that the new pure-gauge term in the new natural basis vanishes as well.

To sum up, the Chen \emph{et al.} approach is similar to the famous Stueckelberg trick. A consequence of this mechanism is that, because of the additional Stueckelberg symmetry, the pure-gauge condition is not sufficient to determine uniquely $A^\text{phys}_\mu$. One therefore needs to impose a further constraint  on $A^\text{phys}_\mu$. Such a constraint breaks explicitly the Stueckelberg symmetry, but preserves nevertheless the original gauge symmetry. In other words, deciding which gauge is the natural one does not break the gauge invariance. We postpone to section \ref{uniqueness} the discussion about the problem of uniqueness in the Chen \emph{et al.} approach.

\section{Lorentz covariance}\label{LorentzSection}

Some people questioned the Lorentz covariance of the decomposition  $A_\mu=A^\text{pure}_\mu+A^\text{phys}_\mu$. The issue is the following: does the physical part remain physical after a Lorentz transformation? To answer this question, one has to determine the Lorentz transformation law of the gauge potential. It is very common and convenient to think of this gauge potential as a Lorentz four-vector. Standard textbooks on classical electrodynamics, like \emph{e.g.} \cite{Jackson,Feynman}, even argue that the gauge potential \emph{must} be a Lorentz four-vector.  On the other hand, standard textbooks on quantum field theory, like \emph{e.g.} \cite{Weinberg,BjorkenDrell}, argue that the gauge potential \emph{cannot} be a Lorentz four-vector. So the situation appears somewhat confusing. 

Note that the conclusion of Refs. \cite{Weinberg,BjorkenDrell} applies actually only to the physical degrees of freedom contained in the gauge potential, and therefore does not proscribe the use of $A_\mu$ as a Lorentz four-vector. On the other hand, we are going to show in this section that the standard argument used in classical electrodynamics textbooks is actually not a proof owing to a loophole. Then, we will show that the Lorentz invariance alone just tells us that, in general, the gauge potential transforms as a Lorentz four-vector \emph{up to} a gauge transformation. So one has the \emph{freedom} to consider it as a Lorentz four-vector, but this is not a necessity. Choosing an inappropriate Lorentz transformation law will generally mix physical and gauge degrees of freedom. But if one chooses the appropriate Lorentz transformation law, the physical part of the gauge potential will remain physical in any Lorentz frame.

\subsection{Loophole in the standard argument}

To stress the Lorentz covariance of the classical laws of electromagnetism and deal with expressions that are simple to transform from one Lorentz frame to another, one tries to reformulate these laws in a \emph{manifestly} Lorentz-covariant form, \emph{i.e.} in terms of Lorentz four-vectors and tensors. Combining the electric and magnetic fields into an antisymmetric matrix $F^{\mu\nu}$ such that $E^i=F^{i0}$ and $B^i=-\frac{1}{2}\epsilon^{ijk}F^{jk}$, one can write the Maxwell's equations and the Lorentz force in a compact form
\begin{align}
\partial_\mu F^{\mu\nu}&=j^\nu,\label{inhomogeneous}\\
\tfrac{1}{2}\,\epsilon^{\mu\nu\alpha\beta}\partial_\nu F_{\alpha\beta}&=0,\label{homogeneous}\\
\frac{\ud\pi^\mu}{\ud\tau}&=\frac{e}{m}\,F^{\mu\nu}\pi_\nu,
\end{align}
where $j^\mu=(\rho,\vec j)$ is a Lorentz four-vector owing to the fact that the electric charge is a Lorentz scalar, $\pi^\mu=(m\gamma,m\gamma\vec\beta)$ is the \emph{kinetic} four-momentum proportional to the rest mass $m$, and $\tau$ is the proper time. Clearly, these equations will be Lorentz covariant if $F^{\mu\nu}$ transforms as a Lorentz tensor\footnote{Many textbooks derive the Lorentz transformation laws of the electric and magnetic fields from the fact that $F^{\mu\nu}$ is a Lorentz tensor. Note that the actual \emph{proof} that the classical laws of electromagnetism are Lorentz covariant \emph{derives} from the Lorentz transformation law of the electric and magnetic fields which has to be established experimentally.}
\begin{equation}\label{EMtensor}
F^{\mu\nu}(x)\mapsto F'^{\mu\nu}(x')=\Lambda^\mu_{\phantom{\mu}\alpha}\,\Lambda^\nu_{\phantom{\mu}\beta}\,F^{\alpha\beta}(x).
\end{equation}
Owing to the homogeneous Maxwell's equation~\eqref{homogeneous}, the electromagnetic tensor $F^{\mu\nu}$ can be expressed in terms of a four-component gauge potential $A^\mu=(\Phi,\vec A)$ as
\begin{equation}
F^{\mu\nu}=\partial^\mu A^\nu-\partial^\nu A^\mu.
\end{equation}
In terms of this gauge potential, the inhomogeneous Maxwell's equation~\eqref{inhomogeneous} reads
\begin{equation}
\partial_\mu\partial^\mu A^\nu-\partial^\nu \partial_\mu A^\mu=j^\nu.
\end{equation}

The standard argument consists in using the gauge freedom \eqref{abelian} to simplify this equation. In the family of gauge potentials satisfying the Lorenz condition
\begin{equation}\label{Lorenz}
\partial_\mu A^\mu=0,
\end{equation}
the inhomogeneous Maxwell's equation reduces to
\begin{equation}\label{Lorenzreduced}
\partial_\mu\partial^\mu A^\nu=j^\nu.
\end{equation}
Since $\partial_\mu\partial^\mu$ is a Lorentz scalar operator and $j^\nu$ is a Lorentz four-vector, the standard conclusion is that $A^\mu$ has to transform as a Lorentz four-vector \cite{Jackson,Feynman}. It is however important to realize that for this to be true, one has actually to make further implicit assumptions. First note that a gauge transformation satisfying $\partial_\mu\partial^\mu\alpha=0$ leaves both \eqref{Lorenz} and \eqref{Lorenzreduced} invariant. This means that one cannot conclude that the only possible Lorentz transformation law for the gauge potential is the four-vector one, unless one removes the residual gauge freedom with \emph{e.g.} some boundary conditions. On top of that, one has also to assume that the Lorenz condition is Lorentz covariant, simply because imposing a non-covariant condition on a covariant equation leads to a non-covariant equation. So, instead of proving that $A_\mu$ transforms as a Lorentz four-vector, one rather implicitly assumes it.

The most general Lorentz transformation law for $A^\mu$ which is consistent with Eq.~\eqref{EMtensor} is actually
\begin{equation}\label{Apot}
A^\mu(x)\mapsto A'^\mu(x')=\Lambda^\mu_{\phantom{\mu}\nu}\left[A^\nu(x)+\partial^\nu\Omega_\Lambda(x)\right],
\end{equation}
where $\Omega_\Lambda$ is a function of space and time associated with the Lorentz transformation $\Lambda$. So in general $A^\mu$ transforms as a Lorentz four-vector only up to a gauge transformation \cite{BjorkenDrell,Weinberg,Manoukian:1987hy,Moriyasu}. The gauge symmetry forbids any determination of the actual function $\Omega_\Lambda$. In gauge theories, there are therefore intrinsically an infinite number of physically equivalent Lorentz transformation laws. The standard argument considers that the Lorenz gauge condition is special. However, it is important to realize that the Lorenz gauge condition has intrinsically nothing special unless one already thinks of $A^\mu$ as a four-vector. In general, one can choose any favorite gauge condition. Thanks to the gauge symmetry, it is possible to impose this gauge condition in any Lorentz frame. Consequently, there exists a subset of Lorentz transformation laws that leave this gauge condition invariant. In this restricted class of Lorentz transformation laws, the favorite gauge condition then appears more ``natural'' than the other ones simply because it is preserved under Lorentz transformations. The standard argument constitutes only one of the possibilities and cannot therefore be considered as a proof.

As a concrete example, consider the Coulomb condition $\vec\nabla\cdot\vec A=0$. It is often said that such a condition is not Lorentz covariant, simply because it cannot be written in a tensorial form. This is actually wrong. The correct statement is that the Coulomb condition is not \emph{manifestly} Lorentz covariant. As discussed in \cite{BjorkenDrell}, performing Lorentz transformations derived with Noether's theorem from the Lagrangian in the Coulomb gauge  leaves the Coulomb condition invariant. Lorentz covariant expressions may look like Lorentz variant when one deals with non-tensorial objects such as the gauge potential.

\subsection{General Lorentz transformation laws}

Let us now discuss the Lorentz transformation properties from the geometrical point of view. A given gauge theory is determined by the choice of both a gauge symmetry group and a representation of the source field $\psi$. The standard representation is the simplest one where Lorentz transformations act only on space-time indices. For example, a spinor carrying an internal-space index is assumed to transform in the standard representation as
\begin{equation}
\psi(x)\mapsto\psi'(x')=S[\Lambda]\psi(x),
\end{equation}
where $S[\Lambda]$ is the standard matrix representing the Lorentz transformation in Dirac space. By construction, the covariant derivative of the source field transforms according to
\begin{equation}
D_\mu\psi(x)\mapsto(D_\mu\psi)'(x')=\Lambda_\mu^{\phantom{\mu}\nu}S[\Lambda]D_\nu\psi(x),
\end{equation}
from which one deduces immediately the Lorentz transformation laws of the covariant derivative and of the connection
\begin{align}
D_\mu&\mapsto D'_\mu=\Lambda_\mu^{\phantom{\mu}\nu}D_\nu,\\
A_\mu(x)&\mapsto A'_\mu(x')=\Lambda_\mu^{\phantom{\mu}\nu}A_\nu(x).
\end{align}
So, in the standard representation $A_\mu$ transforms as a Lorentz four-vector.

As emphasized in the previous subsection, because of the gauge symmetry, there are infinitely many \emph{equivalent} representations, all connected by a gauge transformation. The general non-standard, but physically equivalent, Lorentz transformation law is therefore
\begin{equation}\label{nonstandard}
\psi(x)\mapsto\psi'(x')=U_\Lambda(x)S[\Lambda]\psi(x).
\end{equation}
Despite appearances, Eq.~\eqref{nonstandard} \emph{is not} a Lorentz transformation followed by a gauge transformation. It is \emph{by definition} the Lorentz transformation in the non-standard representation. One then easily obtains the general Lorentz transformation laws of the covariant derivative and of the connection
\begin{align}
D_\mu&\mapsto D'_\mu=\Lambda_\mu^{\phantom{\mu}\nu}U_\Lambda(x)D_\nu U_\Lambda^{-1}(x),\label{covLorentz}\\
A_\mu(x)&\mapsto A'_\mu(x')=\Lambda_\mu^{\phantom{\mu}\nu}U_\Lambda(x)\left[A_\nu(x)+\frac{i}{g}\,\partial_\nu\right]U_\Lambda^{-1}(x).\label{connectionlaw}
\end{align}
The standard representation is naturally recovered using $U_\Lambda(x)=\mathds{1}$. In electrodynamics, one has $g=-e$ and $U_\Lambda(x)=e^{-ie\Omega_\Lambda(x)}$, so that the Lorentz transformation law \eqref{connectionlaw} reduces to \eqref{Apot}. The Lorentz transformation law \eqref{connectionlaw} shows that $A_\mu$ generally transforms as a connection. Indeed, writing explicitly the internal indices
\begin{equation}
-igA^a_{\mu b}\mapsto -igA'^a_{\mu b}=(\Lambda^{-1})_{\phantom{\nu}\mu}^\nu\,(U^{-1}_\Lambda)^d_{\phantom{d}b}\,(U_\Lambda)^a_{\phantom{a}c}\,(-igA^c_{\nu d})+(U_\Lambda)^a_{\phantom{a}e}\left[\partial'_\mu(U^{-1}_\Lambda)^e_{\phantom{e}b}\right],
\end{equation}
one sees that it has exactly the same structure as the Lorentz transformation law of the Christoffel symbols
\begin{equation}
\Gamma^\lambda_{\mu\nu}\mapsto \Gamma'^\lambda_{\mu\nu}=\frac{\partial x^\alpha}{\partial x'^\mu}\,\frac{\partial x^\beta}{\partial x'^\nu}\,\frac{\partial x'^\lambda}{\partial x^\gamma}\,\Gamma^\gamma_{\alpha\beta}+\frac{\partial x'^\lambda}{\partial x^\rho}\,\frac{\partial^2x^\rho}{\partial x'^\mu\partial x'^\nu},
\end{equation}
which is familiar from general relativity, see \emph{e.g.} \cite{Soper}. The corresponding field strength tensor generally transforms as
\begin{equation}
F_{\mu\nu}(x)\mapsto F'_{\mu\nu}(x')=\Lambda_\mu^{\phantom{\mu}\alpha}\Lambda_\nu^{\phantom{\nu}\beta}U_\Lambda(x)F_{\alpha\beta}(x)U^{-1}_\Lambda,
\end{equation}
or more explicitly
\begin{equation}
F^a_{\mu\nu b}\mapsto F'^a_{\mu\nu b}=(\Lambda^{-1})_{\phantom{\alpha}\mu}^\alpha\,(\Lambda^{-1})_{\phantom{\beta}\nu}^\beta\,(U^{-1}_\Lambda)^d_{\phantom{d}b}\,(U_\Lambda)^a_{\phantom{a}c}\,F^c_{\alpha\beta d},
\end{equation}
which is similar to the Lorentz transformation law of the Riemann curvature tensor
\begin{equation}
R^\lambda_{\phantom{\lambda}\mu\nu\rho}\mapsto R'^\lambda_{\phantom{\lambda}\mu\nu\rho}=\frac{\partial x^\alpha}{\partial x'^\mu}\,\frac{\partial x^\beta}{\partial x'^\nu}\,\frac{\partial x^\delta}{\partial x'^\rho}\,\frac{\partial x'^\lambda}{\partial x^\gamma}\,R^\gamma_{\phantom{\gamma}\alpha\beta\delta}.
\end{equation}

\subsection{Lorentz covariance of the Chen \emph{et al.} approach}

As emphasized in standard textbooks on quantum field theory like \emph{e.g.} \cite{Weinberg,BjorkenDrell}, the physical degrees of freedom (\emph{i.e.} the physical part) of the gauge potential in QED cannot form a Lorentz four-vector\footnote{According to the theory of massless representations of the Lorentz group, the only physical massless four-vector is the gradient of a scalar field $\partial^\mu\phi$ and has therefore spin 0.} but necessarily transforms as
\begin{equation}
A_\mu^\text{phys}(x)\mapsto A'^{\text{phys}}_\mu(x')=\Lambda_\mu^{\phantom{\mu}\nu}\,[A_\nu^\text{phys}(x)+\partial_\nu\Omega_\Lambda^\text{phys}(x)]\label{physabL}
\end{equation}
with $\Omega_\Lambda^\text{phys}\neq 0$. Accordingly, the pure-gauge part will generally transform as
\begin{equation}
A_\mu^\text{pure}(x)\mapsto A'^{\text{pure}}_\mu(x')=\Lambda_\mu^{\phantom{\mu}\nu}\,[A_\nu^\text{pure}(x)+\partial_\nu\Omega_\Lambda^\text{pure}(x)]\label{pureabL}
\end{equation}
with $\Omega_\Lambda^\text{pure}=\Omega_\Lambda-\Omega_\Lambda^\text{phys}$. Now, remember that one has the freedom to choose the representation, \emph{i.e.} the function $\Omega_\Lambda$. Two options are particularly interesting:
\begin{enumerate}
\item In the first option, one wants the Lorentz covariance to be \emph{manifest} by working with a gauge potential transforming as a Lorentz four-vector $\Omega_\Lambda=0$. The disadvantage of this option is that the pure-gauge part does not transform as a Lorentz four-vector $\Omega_\Lambda^\text{pure}=-\Omega_\Lambda^\text{phys}$, and so Lorentz transformations will generally mix physical and gauge degrees of freedom. In other words, each time the Lorentz frame is changed, one needs to perform an additional gauge transformation to recover the physical polarizations.
\item In the second option, one requires that physical and gauge degrees do not mix under Lorentz transformations $\Omega_\Lambda^\text{pure}=0$. In other words, physical polarizations remain physical after Lorentz transformations. The disadvantage of this option is that the gauge potential necessarily has a complicated Lorentz transformation law $\Omega_\Lambda=\Omega_\Lambda^\text{phys}$.
\end{enumerate}

In the case of QCD, the situation is analogous, but more complicated owing to the non-abelian nature of the gauge group. In general, the natural components of the source field will transform as
\begin{equation}
\hat\psi(x)\mapsto\hat\psi'(x')=U^\text{phys}_\Lambda(x)S[\Lambda]\hat\psi(x).
\end{equation}
Combining this Lorentz transformation law with \eqref{nonstandard}  and \eqref{connectionlaw} leads to
\begin{align}
U_\text{pure}(x)&\mapsto U'_\text{pure}(x')=U_\Lambda(x)U_\text{pure}(x)U^{\text{phys},-1}_\Lambda(x),\\
A^\text{pure}_\mu(x)&\mapsto A'^{\text{pure}}_\mu(x')=\Lambda_\mu^{\phantom{\mu}\nu}U_\Lambda(x)\left[A_\nu^\text{pure}(x)+\frac{i}{g}\,\partial_\nu\right]U^{-1}_\Lambda(x)\nonumber\\
&\qquad\qquad+\frac{i}{g}\,\Lambda_\mu^{\phantom{\mu}\nu}U_\Lambda(x)U_\text{pure}(x)U^{\text{phys},-1}_\Lambda(x)\left[\partial_\nu U^\text{phys}_\Lambda(x)\right]U^{-1}_\text{pure}(x)U^{-1}_\Lambda(x),\label{pureL}\\
A^\text{phys}_\mu(x)&\mapsto A'^{\text{phys}}_\mu(x')=\Lambda_\mu^{\phantom{\mu}\nu}U_\Lambda(x)A_\nu^\text{phys}(x)U^{-1}_\Lambda(x)\nonumber\\
&\qquad\qquad-\frac{i}{g}\,\Lambda_\mu^{\phantom{\mu}\nu}U_\Lambda(x)U_\text{pure}(x)U^{\text{phys},-1}_\Lambda(x)\left[\partial_\nu U^\text{phys}_\Lambda(x)\right]U^{-1}_\text{pure}(x)U^{-1}_\Lambda(x).\label{physL}
\end{align}
For electrodynamics one has $g=-e$, $U_\Lambda(x)=e^{-ie\Omega_\Lambda(x)}$ and $U^\text{phys}_\Lambda(x)=e^{-ie\Omega^\text{phys}_\Lambda(x)}$, so that the Lorentz transformation laws \eqref{pureL} and \eqref{physL} reduce to the abelian ones \eqref{pureabL} and \eqref{physabL}, respectively. Once again, we have the freedom to choose the representation, \emph{i.e.} the unitary function $U_\Lambda$. Like in the abelian case, two options are particularly interesting:
\begin{enumerate}
\item In the first option, one works with a gauge potential transforming as a Lorentz four-vector $U_\Lambda=\mathds{1}$. Again, with this choice, Lorentz transformations will generally mix physical and gauge degrees of freedom. For example, suppose that in a given Lorentz frame one has chosen to work in the natural gauge, \emph{i.e.} with $U_\text{pure}=\mathds{1}$ and consequently $A^\text{pure}_\mu=0$. After a Lorentz transformation, the pure-gauge part becomes $A'^{\text{pure}}_\mu=\frac{i}{g}\,\Lambda_\mu^{\phantom{\mu}\nu}U^{\text{phys},-1}_\Lambda\partial_\nu U^\text{phys}_\Lambda$. One therefore needs to perform an additional gauge transformation with $U=U^{\text{phys},-1}_\Lambda$ in order to recover a vanishing pure-gauge part $\tilde A'^{\text{pure}}_\mu=0$ in the new Lorentz frame.
\item The second option consists in using $U_\Lambda$ which satisfies the condition $U^{-1}_\Lambda\partial_\mu U_\Lambda=U_\text{pure}U^{\text{phys},-1}_\Lambda\left(\partial_\mu U^\text{phys}_\Lambda\right)U^{-1}_\text{pure}$. In this case, the gauge potential transforms in the same way as its physical part, while the pure-gauge part undergoes a simple rotation in the internal space on top of a four-vector transformation in the physical space $A'^{\text{pure}}_\mu=\Lambda_\mu^{\phantom{\mu}\nu}U_\Lambda A^\text{pure}_\nu U^{-1}_\Lambda$. Consequently, the physical polarizations remain physical under Lorentz transformations, but are generally rotated in the internal space. This internal rotation comes from the fact that different observers may not agree on the ``color'' of a quark. Note that when the observers manage to agree on what is ``red'', ``green'' and ``blue'', we expect $U^\text{phys}_\Lambda$ to reduce to a simple phase factor like in the abelian case.
\end{enumerate}

The first option is so widely adopted in the literature that one has the impression that it is the only acceptable one, and that the gauge invariance is somehow a consequence of the Lorentz covariance \cite{Weinberg}. From this perspective, one would then conclude that the Chen \emph{et al.} approach cannot be Lorentz covariant. However, from the perspective of the second option, gauge invariance and Lorentz covariance can be decoupled. So, with the appropriate Lorentz transformation law for the gauge potential $A_\mu$, one can make the Chen \emph{et al.} approach Lorentz covariant. Note however that it cannot be made \emph{manifestly} Lorentz covariant, in the sense that one is forced to work with objects that do not transform as the usual (and more familiar) Lorentz tensors.

\section{Momentum and angular momentum decompositions}\label{DecompositionSection}

In this section, we remind and compare the main different decompositions of the proton momentum and angular momentum. We follow the covariant point of view adopted by Wakamatsu \cite{Wakamatsu:2010cb} and show that all the total generators are gauge invariant and coincide with each other.

\subsection{Tensor densities}

The QCD Lagrangian is made of three terms $\uL_\text{QCD}=\uL_\text{D}+\uL_\text{YM}+\uL_\text{int}$,
where the so-called Dirac, Yang-Mills and interaction terms are given by
\begin{subequations}
\begin{align}
\uL_\text{D}&=\barpsi(i\gamma^\mu\partial_\mu-m)\psi,\\
\uL_\text{YM}&=-\frac{1}{2}\,\uTr[F^{\mu\nu}F_{\mu\nu}],\\
\uL_\text{int}&=\barpsi\gamma^\mu A_\mu\psi\label{Lint}.
\end{align}
\end{subequations}
For later convenience, the interaction term can be further decomposed into pure-gauge and physical parts $\uL_\text{int}=\uL^\text{pure}_\text{int}+\uL^\text{phys}_\text{int}$ with $\uL^\text{pure}_\text{int}=\barpsi\gamma^\mu A^\text{pure}_\mu\psi$ and $\uL^\text{phys}_\text{int}=\barpsi\gamma^\mu A^\text{phys}_\mu\psi$.

The \emph{canonical} stress-energy and covariant angular momentum tensor densities are obtained directly from Noether's theorem applied to this QCD Lagrangian
\begin{align}
T^{\mu\nu}_\text{c}&=\frac{i}{2}\,\barpsi \gamma^\mu\!\!\stackrel{\leftrightarrow}{\partial}\!\!\!\!\!\phantom{\partial}^\nu\psi-2\,\uTr[F^{\mu\alpha}\partial^\nu A_\alpha]-g^{\mu\nu}\uL_\text{QCD},\\
M^{\mu\nu\rho}_\text{c}&=\frac{1}{2}\,\epsilon^{\mu\nu\rho\sigma}\barpsi\gamma_\sigma\gamma_5\psi+\frac{i}{2}\,\barpsi \gamma^\mu x^{[\nu}\!\!\stackrel{\leftrightarrow}{\partial}\!\!\!\!\!\phantom{\partial}^{\rho]}\psi+\frac{i}{2}\,g^{\mu[\nu}\barpsi\gamma^{\rho]}\psi\nonumber\\
&\qquad-2\,\uTr[F^{\mu[\nu} A^{\rho]}]-2\,\uTr[F^{\mu\alpha}x^{[\nu} \partial^{\rho]}A_\alpha]-x^{[\nu}g^{\rho]\mu}\uL_\text{QCD},
\end{align}
where $\stackrel{\leftrightarrow}{\partial}=\stackrel{\rightarrow}{\partial}-\stackrel{\leftarrow}{\partial}$, $a^{[\mu}b^{\nu]}=a^\mu b^\nu-a^\nu b^\mu$, and $\epsilon_{0123}=1$. According to Noether's theorem, both $T^{\mu\nu}_\text{c}$ and $M^{\mu\nu\rho}_\text{c}$ are conserved $\partial_\mu T^{\mu\nu}_\text{c}=\partial_\mu M^{\mu\nu\rho}_\text{c}=0$. As one can easily see, the problem with these densities is that they are not gauge invariant
\begin{equation}
\begin{split}
T^{\mu\nu}_\text{c}&\mapsto\tilde T^{\mu\nu}_\text{c}=T^{\mu\nu}_\text{c}-\frac{2i}{g} \,\partial_\alpha\uTr[F^{\mu\alpha}(\partial^\nu U^{-1})U],\\
M^{\mu\nu\rho}_\text{c}&\mapsto\tilde M^{\mu\nu\rho}_\text{c}=M^{\mu\nu\rho}_\text{c}-\frac{2i}{g} \,\partial_\alpha\uTr[F^{\mu\alpha}x^{[\nu}(\partial^{\rho]}U^{-1})U].
\end{split}
\end{equation}

Alternatively, one can consider the following manifestly gauge-invariant tensor densities
\begin{align}
\mathcal T^{\mu\nu}_\text{gi}&=\frac{i}{2}\,\barpsi \gamma^\mu\!\!\stackrel{\leftrightarrow}{D}\!\!\!\!\!\phantom{\partial}^\nu\psi-2\,\uTr[F^{\mu\alpha}F^\nu_{\phantom{\nu}\alpha}]-g^{\mu\nu}\uL_\text{QCD},\\
\mathcal M^{\mu\nu\rho}_\text{gi}&=\frac{1}{2}\,\epsilon^{\mu\nu\rho\sigma}\barpsi\gamma_\sigma\gamma_5\psi+\frac{i}{2}\,\barpsi \gamma^\mu x^{[\nu}\!\!\stackrel{\leftrightarrow}{D}\!\!\!\!\!\phantom{\partial}^{\rho]}\psi-2\,\uTr[F^{\mu\alpha}x^{[\nu}F^{\rho]}_{\phantom{\rho}\alpha}]-x^{[\nu}g^{\rho]\mu}\uL_\text{QCD}.
\end{align}
They differ from the canonical ones just by a four-divergence term\footnote{These four-divergence terms are called superpotential in Ref. \cite{Jaffe:1989jz}.}
\begin{equation}
\begin{split}
\mathcal T^{\mu\nu}_\text{gi}&=T^{\mu\nu}_\text{c}+2\partial_\alpha\uTr[F^{\mu\alpha}A^\nu],\\
\mathcal M^{\mu\nu\rho}_\text{gi}&=M^{\mu\nu\rho}_\text{c}+2\partial_\alpha\uTr[F^{\mu\alpha}x^{[\nu}A^{\rho]}].
\end{split}
\end{equation}
Since the field strength tensor is antisymmetric, the conservation of the canonical tensor densities implies the conservation of the gauge-invariant ones.

In the Chen \emph{et al.} approach, one considers instead other manifestly gauge-invariant tensor densities
\begin{align}
\mathsf T^{\mu\nu}_\text{Chen}&=\frac{i}{2}\,\barpsi \gamma^\mu\!\!\stackrel{\leftrightarrow}{D}\!\!\!\!\!\phantom{\partial}^\nu_\text{pure}\psi-2\,\uTr[F^{\mu\alpha}\mathcal D^\nu_\text{pure} A^\text{phys}_\alpha]-g^{\mu\nu}\uL_\text{QCD},\\
\mathsf M^{\mu\nu\rho}_\text{Chen}&=\frac{1}{2}\,\epsilon^{\mu\nu\rho\sigma}\barpsi\gamma_\sigma\gamma_5\psi+\frac{i}{2}\,\barpsi \gamma^\mu x^{[\nu}\!\!\stackrel{\leftrightarrow}{D}\!\!\!\!\!\phantom{\partial}^{\rho]}_\text{pure}\psi+\frac{i}{2}\,g^{\mu[\nu}\barpsi\gamma^{\rho]}\psi\nonumber\\
&\qquad-2\,\uTr[F^{\mu[\nu} A^{\rho]}_\text{phys}]-2\,\uTr[F^{\mu\alpha}x^{[\nu} \mathcal D^{\rho]}_\text{pure}A^\text{phys}_\alpha]-x^{[\nu}g^{\rho]\mu}\uL_\text{QCD}.
\end{align}
They also differ from the canonical ones just by a four-divergence term
\begin{equation}
\begin{split}
\mathsf T^{\mu\nu}_\text{Chen}&=T^{\mu\nu}_\text{c}+2\partial_\alpha\uTr[F^{\mu\alpha}A^\nu_\text{pure}],\\
\mathsf M^{\mu\nu\rho}_\text{Chen}&=M^{\mu\nu\rho}_\text{c}+2\partial_\alpha\uTr[F^{\mu\alpha}x^{[\nu}A^{\rho]}_\text{pure}],
\end{split}
\end{equation}
and are obviously also conserved. Clearly, the two sets of gauge-invariant tensor densities are simply related by a gauge-invariant four-divergence term
\begin{equation}
\begin{split}
\mathcal T^{\mu\nu}_\text{gi}&=\mathsf T^{\mu\nu}_\text{Chen}+2\partial_\alpha\uTr[F^{\mu\alpha}A^\nu_\text{phys}],\\
\mathcal M^{\mu\nu\rho}_\text{gi}&=\mathsf M^{\mu\nu\rho}_\text{Chen}+2\partial_\alpha\uTr[F^{\mu\alpha}x^{[\nu}A^{\rho]}_\text{phys}].
\end{split}
\end{equation}

Provided that surface terms vanish\footnote{Note that this might not be justified in QCD because of the Gribov ambiguities for the non-perturbative non-abelian gauge field configuration, and because gluon-field configurations with non-trivial topology might play some role in the nucleon structure.}, one sees that the three different sets of conserved tensor densities give the \emph{same} set of time-independent charges
\begin{equation}\label{charges}
\begin{split}
 P^\nu&=\int\ud^3x\,n_\mu T^{\mu\nu}_\text{c}=\int\ud^3x\,n_\mu\mathcal T^{\mu\nu}_\text{gi}=\int\ud^3x\,n_\mu\mathsf T^{\mu\nu}_\text{Chen},\\
 J^{\nu\rho}&=\int\ud^3x\,n_\mu M^{\mu\nu\rho}_\text{c}=\int\ud^3x\,n_\mu\mathcal M^{\mu\nu\rho}_\text{gi}=\int\ud^3x\,n_\mu\mathsf M^{\mu\nu\rho}_\text{Chen},
\end{split}
\end{equation}
where $n_\mu=(1,0,0,0)$ in the instant form and $n_\mu=(1,0,0,1)/\sqrt{2}$ in the light-front form. According to Noether's theorem, since these charges are obtained from the canonical densities, they are \emph{total} generators of space-time translations and Lorentz transformations, and are consequently identified with the total energy-momentum and four-angular momentum operators. In a recent paper \cite{Leader:2011za}, Leader proposed a proof that, in covariantly quantized quantum electrodynamics, the total generators of space-time translations and rotations cannot be gauge-invariant operators. On the other hand, Eqs.~\eqref{charges} show that the (classical) total generators can explicitly be expressed in terms of gauge-invariant quantities only. Further investigations are therefore needed to clarify this point.

\subsection{Gauge-variant and invariant decompositions}

We are interested in how these densities receive contributions from quarks and gluons
\begin{equation}
T^{\mu\nu}=T^{\mu\nu}_q+T^{\mu\nu}_g,\qquad
M^{\mu\nu\rho}=M^{\mu\nu\rho}_q+M^{\mu\nu\rho}_g.
\end{equation}
We are also interested in how the covariant angular momentum tensor density receives contribution from spin and orbital angular momentum (OAM)
\begin{equation}
M^{\mu\nu\rho}=M^{\mu\nu\rho}_\text{spin}+M^{\mu\nu\rho}_\text{OAM}+M^{\mu\nu\rho}_\text{boost},
\end{equation}
where $M^{\mu\nu\rho}_\text{boost}$ contributes only to Lorentz boosts, and so has nothing to do with the nucleon momentum and spin decompositions. As stressed by Leader \cite{Leader:2011za}, it is important to remember that, contrary to the total densities, the individual contributions are not conserved, and consequently the corresponding charges are time dependent. However, their matrix elements are time independent as long as one considers states with a given energy.

According to Jaffe and Manohar \cite{Jaffe:1989jz}, one should use the canonical tensor densities and identify the quark and gluon contributions with their Dirac and pure Yang-Mills expressions
\begin{subequations}\label{JMdecomposition}
\begin{align}
T^{\mu\nu}_q&=\frac{i}{2}\,\barpsi \gamma^\mu\!\!\stackrel{\leftrightarrow}{\partial}\!\!\!\!\!\phantom{\partial}^\nu\psi-g^{\mu\nu}\uL_\text{D},\\
T^{\mu\nu}_g&=-2\,\uTr[F^{\mu\alpha}\partial^\nu A_\alpha]-g^{\mu\nu}\left(\uL_\text{YM}+\uL_\text{int}\right),\\
M^{\mu\nu\rho}_{q,\text{spin}}&=\frac{1}{2}\,\epsilon^{\mu\nu\rho\sigma}\barpsi\gamma_\sigma\gamma_5\psi,\\
M^{\mu\nu\rho}_{q,\text{OAM}}&=\frac{i}{2}\,\barpsi \gamma^\mu x^{[\nu}\!\!\stackrel{\leftrightarrow}{\partial}\!\!\!\!\!\phantom{\partial}^{\rho]}\psi,\\
M^{\mu\nu\rho}_{q,\text{boost}}&=\frac{i}{2}\,g^{\mu[\nu}\barpsi\gamma^{\rho]}\psi-x^{[\nu}g^{\rho]\mu}\uL_\text{D},\\
M^{\mu\nu\rho}_{g,\text{spin}}&=-2\,\uTr[F^{\mu[\nu} A^{\rho]}],\\
M^{\mu\nu\rho}_{g,\text{OAM}}&=-2\,\uTr[F^{\mu\alpha}x^{[\nu} \partial^{\rho]}A_\alpha],\\
M^{\mu\nu\rho}_{g,\text{boost}}&=-x^{[\nu}g^{\rho]\mu}\left(\uL_\text{YM}+\uL_\text{int}\right).
\end{align}
\end{subequations}
The problem with such a decomposition is that, except for $M^{\mu\nu\rho}_{q,\text{spin}}$, the contributions are not gauge invariant, making their physical meaning questionable. Such a decomposition is then meaningful only when the gauge is fixed. The standard choice is $A^+=0$ in order to make contact with the parton model picture \cite{Jaffe:1989jz,Bashinsky:1998if}. 

To cure this problem, Ji \cite{Ji:1996ek} proposed to use instead the gauge-invariant tensor densities and to decompose them in the following way
\begin{subequations}
\begin{align}
\mathcal T^{\mu\nu}_q&=\frac{i}{2}\,\barpsi \gamma^\mu\!\!\stackrel{\leftrightarrow}{D}\!\!\!\!\!\phantom{\partial}^\nu\psi-g^{\mu\nu}\left(\uL_\text{D}+\uL_\text{int}\right),\\
\mathcal T^{\mu\nu}_g&=-2\,\uTr[F^{\mu\alpha}F^\nu_{\phantom{\nu}\alpha}]-g^{\mu\nu}\uL_\text{YM},\\
\mathcal M^{\mu\nu\rho}_{q,\text{spin}}&=\frac{1}{2}\,\epsilon^{\mu\nu\rho\sigma}\barpsi\gamma_\sigma\gamma_5\psi,\\
\mathcal M^{\mu\nu\rho}_{q,\text{OAM}}&=\frac{i}{2}\,\barpsi \gamma^\mu x^{[\nu}\!\!\stackrel{\leftrightarrow}{D}\!\!\!\!\!\phantom{\partial}^{\rho]}\psi,\\
\mathcal M^{\mu\nu\rho}_{q,\text{boost}}&=\frac{i}{2}\,g^{\mu[\nu}\barpsi\gamma^{\rho]}\psi-x^{[\nu}g^{\rho]\mu}\left(\uL_\text{D}+\uL_\text{int}\right),\\
\mathcal M^{\mu\nu\rho}_{g,\text{spin}+\text{OAM}}&=-2\,\uTr[F^{\mu\alpha}x^{[\nu}F^{\rho]}_{\phantom{\rho}\alpha}],\\
\mathcal M^{\mu\nu\rho}_{g,\text{boost}}&=-x^{[\nu}g^{\rho]\mu}\uL_\text{YM},
\end{align}
\end{subequations}
where each term is obviously gauge invariant. As mentioned in the introduction, there is however no further decomposition of the gluon angular momentum into spin and OAM.

The Chen \emph{et al.} approach is partly motivated by the ability to separate the gluon angular momentum into spin and OAM in a gauge-invariant way. In this case, one uses the alternative gauge-invariant tensor densities and decompose them in the following way
\begin{subequations}\label{Chenetaldec}
\begin{align}
\mathsf T^{\mu\nu}_q&=\frac{i}{2}\,\barpsi \gamma^\mu\!\!\stackrel{\leftrightarrow}{D}\!\!\!\!\!\phantom{\partial}^\nu_\text{pure}\psi-g^{\mu\nu}\left(\uL_\text{D}+\uL^\text{pure}_\text{int}\right),\\
\mathsf T^{\mu\nu}_g&=-2\,\uTr[F^{\mu\alpha}\mathcal D^\nu_\text{pure} A^\text{phys}_\alpha]-g^{\mu\nu}\left(\uL_\text{YM}+\uL^\text{phys}_\text{int}\right),\\
\mathsf M^{\mu\nu\rho}_{q,\text{spin}}&=\frac{1}{2}\,\epsilon^{\mu\nu\rho\sigma}\barpsi\gamma_\sigma\gamma_5\psi,\\
\mathsf M^{\mu\nu\rho}_{q,\text{OAM}}&=\frac{i}{2}\,\barpsi \gamma^\mu x^{[\nu}\!\!\stackrel{\leftrightarrow}{D}\!\!\!\!\!\phantom{\partial}^{\rho]}_\text{pure}\psi,\\
\mathsf M^{\mu\nu\rho}_{q,\text{boost}}&=\frac{i}{2}\,g^{\mu[\nu}\barpsi\gamma^{\rho]}\psi-x^{[\nu}g^{\rho]\mu}\left(\uL_\text{D}+\uL^\text{pure}_\text{int}\right),\\
\mathsf M^{\mu\nu\rho}_{g,\text{spin}}&=-2\,\uTr[F^{\mu[\nu} A^{\rho]}_\text{phys}],\label{Gspin}\\
\mathsf M^{\mu\nu\rho}_{g,\text{OAM}}&=-2\,\uTr[F^{\mu\alpha}x^{[\nu} \mathcal D^{\rho]}_\text{pure}A^\text{phys}_\alpha],\\
\mathsf M^{\mu\nu\rho}_{g,\text{boost}}&=-x^{[\nu}g^{\rho]\mu}\left(\uL_\text{YM}+\uL^\text{phys}_\text{int}\right),
\end{align}
\end{subequations}
where each term is also obviously gauge invariant. This gauge-invariant decomposition has a strong resemblance with the Jaffe-Manohar decomposition. When expressed in the natural gauge, \emph{i.e.} when $\tilde A^\mu_\text{pure}=0$ and $\tilde A^\mu=\tilde A^\mu_\text{phys}$, they even become identical. In this sense, it can be thought as a \emph{gauge-invariant extension} (GIE) of the Jaffe-Manohar decomposition  \cite{Wakamatsu:2010cb,Ji:2012gc}.  The concept of GIE can be applied to any gauge-variant quantity like \emph{e.g.} the Chern-Simons current \cite{Guo:2012wv}. What is puzzling about this decomposition is the presence of the pure-gauge covariant derivatives instead of the ordinary ones, which does not suit well with our understanding of classical electrodynamics and the Lorentz force. 

Finally, Wakamatsu \cite{Wakamatsu:2010qj} proposed somehow a compromise between the Ji decomposition  and the GIE. He showed that there exist gauge-invariant terms which can be written, using the QCD equation of motion $\mathcal D^\alpha F^a_{\alpha\mu b}=-g\barpsi_b\gamma_\mu\psi^a$, either as a quark or as a gluon contribution
\begin{align}
\begin{split}
\mathsf T^{\mu\nu}_\text{pot}&=-g\barpsi\gamma^\mu A_\text{phys}^\nu\psi+g^{\mu\nu}\uL^\text{phys}_\text{int}\\
&=2\,\uTr[F^{\mu\alpha}\mathcal D_\alpha A^\nu_\text{phys}]-2\partial_\alpha\uTr[F^{\mu\alpha}A^\nu_\text{phys}]+g^{\mu\nu}\uL^\text{phys}_\text{int},
\end{split}\label{potT}\\
\begin{split}
\mathsf M^{\mu\nu\rho}_\text{pot}&=-g\barpsi\gamma^\mu x^{[\nu}A_\text{phys}^{\rho]}\psi+x^{[\nu}g^{\rho]\mu}\uL^\text{phys}_\text{int}\\
&=2\,\uTr[F^{\mu\alpha} \mathcal D_\alpha( x^{[\nu}A^{\rho]}_\text{phys})]-2\partial_\alpha\uTr[F^{\mu\alpha} x^{[\nu}A^{\rho]}_\text{phys}]+x^{[\nu}g^{\rho]\mu}\uL^\text{phys}_\text{int}.
\end{split}
\end{align} 
They are respectively called \emph{potential momentum} and \emph{potential angular momentum}, following Konopinski's terminology \cite{Konopinski}. In the Ji decomposition, these potential terms are attributed to the gluons, while in the Chen \emph{et al.} approach they are attributed to the quarks
\begin{subequations}
\begin{align}
\mathcal T^{\mu\nu}_q&=\mathsf T^{\mu\nu}_q-\mathsf T^{\mu\nu}_\text{pot},\\
\mathcal T^{\mu\nu}_g&=\mathsf T^{\mu\nu}_g+\mathsf T^{\mu\nu}_\text{pot}+2\partial_\alpha\uTr[F^{\mu\alpha}A^\nu_\text{phys}],\\
\mathcal M^{\mu\nu\rho}_{q,\text{OAM}}&=\mathsf M^{\mu\nu\rho}_{q,\text{OAM}}-\mathsf M^{\mu\nu\rho}_\text{pot},\\
\mathcal M^{\mu\nu\rho}_{g,\text{spin}+\text{OAM}}&=\mathsf M^{\mu\nu\rho}_{g,\text{spin}}+\mathsf M^{\mu\nu\rho}_{g,\text{OAM}}+\mathsf M^{\mu\nu\rho}_\text{pot}+2\partial_\alpha\uTr[F^{\mu\alpha} x^{[\nu}A^{\rho]}_\text{phys}].
\end{align}
\end{subequations}
Wakamatsu argues that the potential terms should be attributed to the gluons, and therefore favors the Ji decomposition \cite{Wakamatsu:2010qj,Wakamatsu:2010cb,Wakamatsu:2011mb,Wakamatsu:2012ve}. Taking advantage of the Chen \emph{et al.} approach, he proposed the following gauge-invariant separation of Ji's gluon angular momentum $\mathcal M^{\mu\nu\rho}_{g,\text{spin}+\text{OAM}}=\mathcal M^{\mu\nu\rho}_{g,\text{spin}}+\mathcal M^{\mu\nu\rho}_{g,\text{OAM}}$ where
\begin{align}
\mathcal M^{\mu\nu\rho}_{g,\text{spin}}&=\mathsf M^{\mu\nu\rho}_{g,\text{spin}},\\
\mathcal M^{\mu\nu\rho}_{g,\text{OAM}}&=\mathsf M^{\mu\nu\rho}_{g,\text{OAM}}+\mathsf M^{\mu\nu\rho}_\text{pot}+2\partial_\alpha\uTr[F^{\mu\alpha} x^{[\nu}A^{\rho]}_\text{phys}].
\end{align}
Note that attributing the potential terms to the quarks is closer to the concept of ``action at a distance'', while attributing it to the gluons is closer to the concept of ``action through a medium''.

\section{Which decomposition to use?}\label{DiscussionSection}

The present situation appears to be quite confusing, particularly because of the number of decompositions that have been proposed. So far, no consensus on which decomposition to use has emerged. Such a consensus might even never be reached. But this is not such a big issue since the controversies mainly concern the physical interpretation. In some sense, the present situation is similar to the early days of quantum mechanics, where people were debating about the interpretation of the theory. Even though the Copenhague interpretation eventually emerged as the dominant one, the debates are still going on. Nervertheless this did not prevent physicists from developing and using quantum mechanics. It would therefore be very useful to come up with a pragmatic approach, leaving aside the ontological questions.  

In this section, we remind the main arguments in favor of the kinetic and the canonical decompositions and show that they rely on two different conceptions of what is the physical momentum. Then we discuss in more detail how the gauge invariance and the canonical operators are reconciled in the Chen \emph{et al.} approach, and how the mechanism \`{a} la Stueckelberg converts a problem of gauge invariance into a problem of uniqueness. We propose a pragmatic point of view and discuss the observability of the OAM.

\subsection{Kinetic \emph{versus} canonical}

As summarized by Wakamatsu \cite{Wakamatsu:2010cb}, all the proposed decompositions can be sorted into essentially two families\footnote{These families are called decompositions (I) and (II) in Wakamatsu's terminology. This classification has been criticized by Leader \cite{Leader:2011za} based on the observation that one can actually consider an infinite number of families by attributing a fraction $\alpha$ of the potential terms to the quarks and the remaining fraction $(1-\alpha)$ to the gluons. Note however that no decompositions with $\alpha\neq 0,1$ have been proposed so far. The reason is that they appear to be quite unnatural as the corresponding momenta and orbital angular momenta are neither kinetic nor canonical.}: 
\begin{itemize}
\item In the \emph{kinetic} family, the potential terms are attributed to the gluons, and so only ordinary covariant derivatives are involved. This family contains therefore the Ji decomposition  and Wakamatsu's improvement. 
\item In the \emph{canonical} family, the potential terms are attributed to the quarks, and so only pure-gauge covariant derivatives are involved. This family contains therefore the Jaffe-Manohar decomposition  and the GIEs.
\end{itemize}
Since the potential terms give non-vanishing physical results, decompositions belonging to different families are necessarily physically inequivalent. While the difference is small in non-relativistic systems like the atoms \cite{Burkardt:2008ua,Wakamatsu:2012ve,Ji:2012gc}, it becomes significant for relativistic systems like the proton \cite{Chen:2009mr,Cho:2011ee}. Deciding which family is the most physical one is at the heart of ongoing debates.

Already at the classical level, there exist two kinds of momentum. One is the kinetic momentum defined as $\vec\pi=m\vec v$, where $m$ and $\vec v=\ud\vec x/\ud t$ are the mass and the velocity of the particle, respectively. It corresponds to our classical intuition where particles follow well-defined trajectories. It is also the momentum appearing in the non-relativistic expression for the particle kinetic energy $\vec\pi^2/2m$. The other is the canonical momentum defined as $\vec p=\partial L/\partial\vec v$, where $L$ is the Lagrangian of the system. Like the particle position $\vec x$, it is a dynamical variable in the Hamiltonian formalism. It is also the generator of translations. 

In absence of electromagnetic fields, these two kinds of momentum are usually identified. But the distinction becomes necessary in presence of electromagnetic fields. As an example, let us consider the classical problem of a charged particle moving in a homogeneous \emph{external} magnetic field. By ``external'' we mean that no dynamics is associated with it, so that the total momentum is simply identified with the particle canonical momentum. From translation invariance, it follows that the canonical momentum is conserved, \emph{i.e.} $\vec p$ is time independent. On the other hand, the Lorentz force tells us that the particle follows a helicoidal trajectory. At each instant, the particle kinetic momentum $\vec\pi(t)$ points toward a different direction and is therefore not conserved. The difference between canonical and kinetic momentum $\vec p-\vec\pi(t)=e\int\ud^3x\,\vec A(\vec x,t)$, which is nothing else than the potential momentum \eqref{potT}, can then be interpreted as the kinetic momentum carried by the external electromagnetic field. In a similar way, the famous Feynman paradox of classical electrodynamics \cite{Feynman} shows that the electromagnetic field carries also some kinetic angular momentum.

In a classical picture, it is more natural to consider that the kinetic momentum and angular momentum are the physical ones. The reason is that they have a direct connection with the particle motion in an external field. Moreover, one can always formulate the problems of classical electrodynamics in the Newtonian formalism, and therefore avoid the use of canonical quantities as well as the problem of gauge invariance. In a quantum-mechanical picture, the canonical momentum and angular momentum appear more natural. The reason is that, in absence of well-defined trajectories, the only natural definition of momentum and angular momentum is as the generators of translations and rotations. Moreover, the canonical quantization rules are formulated in the Hamiltonian formalism, and so one can hardly avoid the use of canonical quantities. Depending on the adopted picture, the opinion about which family of decompositions is the physical one will naturally differ.

\subsection{Gauge invariance and commutation relations}

It is clear that the physical quantities have to be gauge invariant. As stressed by Leader \cite{Leader:2011za}, the argument of gauge invariance applies at the level of the matrix elements
\begin{equation}
\langle\tilde\Phi|\tilde O|\tilde\Phi\rangle=\langle\Phi|O|\Phi\rangle.
\end{equation}
One is therefore allowed to work with gauge-variant operators $\tilde O\neq O$ as long as the Hilbert states are also gauge variant $|\tilde\Phi\rangle\neq|\Phi\rangle$. It is however more common to work with gauge-invariant states and therefore with gauge-invariant operators\footnote{Leader considers another possibility \cite{Anselmino:1994gn,Leader:2011za}: provided that the gauge variation of the operator leads to vanishing matrix elements between physical states $\langle\Phi|\tilde O-O|\Phi\rangle=0$, one can use gauge-variant operators together with gauge-invariant states. Note however that this approach does not solve the problem of gauge invariance. It just \emph{hides} this problem inside a more complicated structure for the Hilbert space, and is consequently not really appropriate for the discussion.}. From this perspective, the physical momentum and angular momentum seem to be the kinetic ones.

We have seen from Eq.~\eqref{charges} that, as long as one considers total momentum and angular momentum, there is no difference between kinetic and canonical operators. The difference appears only when one tries to determine the contributions coming from the different constituents of the system. While Ji and Wakamatsu adopt the classical point of view, Chen \emph{et al.} adopt the quantum-mechanical point of view and therefore require that the physical momentum and angular momentum operators have to satisfy the corresponding canonical commutation relations\footnote{Since the commutation relations are not preserved under renormalization \cite{Shore:1999be,Ji:2010zza}, this requirement is meant only for the \emph{bare} operators \cite{Chen:2008ja}.} \cite{Chen:2008ja,Chen:2009dg}
\begin{equation}\label{comrel}
[P^i_X,P^j_X]=0,\qquad [J^i_X,J^j_X]=i\epsilon^{ijk}J^k_X,\qquad [J^i_X,P^j_X]=i\epsilon^{ijk}P^k_X,
\end{equation}
where $X$ stands for the particle species. The kinetic operators are gauge invariant but do not satisfy the canonical commutation relations. On the other hand, the canonical operators satisfy the canonical commutation relations but are not gauge invariant. To solve this problem, Chen \emph{et al.} proposed to separate the gauge potential into pure-gauge and physical parts, and naturally obtained a decomposition with the structure of a GIE \eqref{Chenetaldec}. They checked in particular that each term satisfies Eq.~\eqref{comrel} as required. 

Adopting also the quantum-mechanical point of view, Leader proposed instead the stronger constraint that the physical momentum and angular momentum operators have to be the generators of translations and rotations for the corresponding particles \cite{Leader:2011za}
\begin{align}
[\vec P_X(t),\phi_X(\vec x,t)]&=-i\vec\nabla\phi_X(\vec x,t),\\
[\vec J_X(t),\phi_X(\vec x,t)]&=[\vec x\times(-i\vec\nabla)+\vec S_X]\,\phi_X(\vec x,t),
\end{align}
where $\vec S_X$ is a spin matrix. This automatically implies that the canonical commutation relations \eqref{comrel} are satisfied. Considering as usual the canonical variables $\phi_X=\psi,A_\mu$ he arrived at the conclusion that the only physical operators are the canonical ones. Note that when gauge-variant dynamical variables are used, it should not be surprising to find that the corresponding momentum and angular momentum operators are not gauge invariant. Recently, Chen observed that the gauge-invariant operators appearing in the GIE are actually the generators of translations and rotations for the gauge-invariant fields $\phi_X=\hat\psi,\hat A^\text{phys}_\mu$ \cite{Chen:2012vg}. The reason is simple: in the Chen \emph{et al.} approach, every expression can be rewritten in terms of the natural components only. For example, the GIE becomes
\begin{subequations}
\begin{align}
\mathsf T^{\mu\nu}_q&=\frac{i}{2}\,\hat\barpsi \gamma^\mu\!\!\stackrel{\leftrightarrow}{\partial}\!\!\!\!\!\phantom{\partial}^\nu\hat\psi-g^{\mu\nu}\hat\uL_\text{D},\\
\mathsf T^{\mu\nu}_g&=-2\,\uTr[\hat F^{\mu\alpha}\partial^\nu\hat A^\text{phys}_\alpha]-g^{\mu\nu}\left(\hat\uL_\text{YM}+\hat\uL^\text{phys}_\text{int}\right),\\
\mathsf M^{\mu\nu\rho}_{q,\text{spin}}&=\frac{1}{2}\,\epsilon^{\mu\nu\rho\sigma}\hat\barpsi\gamma_\sigma\gamma_5\hat\psi,\\
\mathsf M^{\mu\nu\rho}_{q,\text{OAM}}&=\frac{i}{2}\,\hat\barpsi \gamma^\mu x^{[\nu}\!\!\stackrel{\leftrightarrow}{\partial}\!\!\!\!\!\phantom{\partial}^{\rho]}\hat\psi,\\
\mathsf M^{\mu\nu\rho}_{q,\text{boost}}&=\frac{i}{2}\,g^{\mu[\nu}\hat\barpsi\gamma^{\rho]}\hat\psi-x^{[\nu}g^{\rho]\mu}\hat\uL_\text{D},\\
\mathsf M^{\mu\nu\rho}_{g,\text{spin}}&=-2\,\uTr[\hat F^{\mu[\nu}\hat A^{\rho]}_\text{phys}],\\
\mathsf M^{\mu\nu\rho}_{g,\text{OAM}}&=-2\,\uTr[\hat F^{\mu\alpha}x^{[\nu} \partial^{\rho]}\hat A^\text{phys}_\alpha],\\
\mathsf M^{\mu\nu\rho}_{g,\text{boost}}&=-x^{[\nu}g^{\rho]\mu}\left(\hat\uL_\text{YM}+\hat\uL^\text{phys}_\text{int}\right),
\end{align}
\end{subequations}
which mimics perfectly the Jaffe-Manohar decomposition \eqref{JMdecomposition}, except that the fields involved are now gauge invariant. In conclusion, the ordinary canonical momentum and angular momentum operators are gauge variant simply because one usually considers gauge-variant fields as canonical variables. Choosing instead gauge-invariant fields as canonical variables leads naturally to gauge-invariant canonical momentum and angular momentum operators.

\subsection{Issues of uniqueness and locality in the Chen \emph{et al.} approach}\label{uniqueness}

We have seen that the Chen \emph{et al.} approach solved the problem of gauge invariance at the price of introducing an additional Stueckelberg symmetry which is then broken by the choice a natural gauge. While the GIE is manifestly gauge invariant, it is not Stueckelberg invariant. In other words, to each choice of a natural gauge corresponds a GIE. While the gauge symmetry implies that the physical quantities have to be gauge invariant, the Stueckelberg symmetry implies that there are infinitely many gauge-invariant canonical momenta and angular momenta\footnote{Ji interprets the existence of an infinity of canonical quantities as a signal that the Chen \emph{et al.} approach is ``not really gauge invariant in the textbook sense'' \cite{Ji:2010zza,Ji:2009fu,Ji:2012gc}. Such a formulation is unfortunate as it gives the impression that, according to Ji, the Chen \emph{et al.} approach is not gauge invariant \cite{Chen:2008ja,Chen:2009dg}. To avoid misunderstandings, it is important to clearly distinguish in the discussions the gauge symmetry from the Stueckelberg symmetry.}. Note that already in Ref. \cite{Bashinsky:1998if} Bashinsky and Jaffe stressed that ``one should make clear what a quark or a gluon parton is in an interacting theory. The subtlety here is the issue of gauge invariance: a pure quark field in one gauge is a superposition of quarks and gluons in another. Different ways of gluon-field gauge fixing predetermine different decompositions of the coupled quark-gluon fields into quark and gluon degrees of freedom. Similarly, one can generalize a gauge-variant non-local operator [\ldots] to more than one gauge-invariant expressions, raising the problem of deciding which is the \emph{true} one.'' 

Since all the GIEs reduce to the Jaffe-Manohar decomposition  in the appropriate gauges, Wakamatsu concluded that they are all physically equivalent \cite{Wakamatsu:2010cb}. So, instead of two families of decompositions, he claims that there are actually only two physically inequivalent decompositions, the kinetic and the canonical ones. Wakamatsu explains that his conclusion is a consequence of gauge invariance. But we have seen that the gauge invariance has nothing to do with the actual choice of $\hat A_\mu^\text{phys}$. In reality, Wakamatsu's conclusion is a consequence of the \emph{assumption} that physical quantities are Stueckelberg invariant. Indeed, expressed in other words, Wakamatsu's claim is that the matrix elements of Stueckelberg-variant operators are Stueckelberg invariant. Since a GIE is after all a gauge-fixed result extended in a gauge-invariant way \cite{Ji:2012gc}, Wakamatsu's claim is simply equivalent to Leader's claim that the matrix elements of gauge-variant operators are gauge invariant \cite{Leader:2011za}.

\begin{figure}[t!]
	\centering
		\includegraphics[width=.5\textwidth]{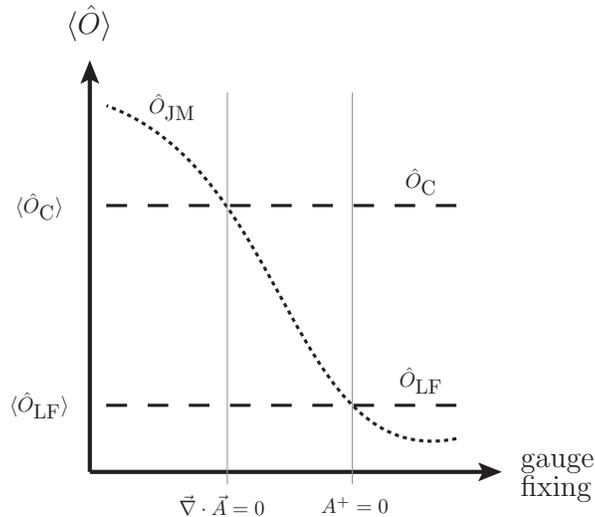}
\caption{\footnotesize{While the operators $\hat O_\text{JM}$ belonging to the Jaffe-Manohar decomposition  lead to different results in different gauges, the operators $\hat O_\text{C}$ and $\hat O_\text{LF}$ belonging respectively to the Coulomb and light-front GIEs are gauge invariant. Since one has $\hat O_\text{C}=\hat O_\text{JM}$ in the Coulomb gauge and $\hat O_\text{LF}=\hat O_\text{JM}$ in the light-front gauge, different GIEs are physically inequivalent $\hat O_\text{C}\neq\hat O_\text{LF}$.}}
		\label{fig1}
\end{figure}

It is however more natural to expect that different GIEs are physically inequivalent. Since Wakamatsu's improvement relies on the Chen \emph{et al.} approach, one would also expect infinitely many decompositions of the kinetic gluon angular momentum into spin and OAM contributions. Trying to determine which GIE is the physical one amounts to decide which gauge is the natural one. To clearly separate the physical degrees of freedom, the natural gauge has to belong to the class of \emph{physical} gauges. There are two popular physical gauges, namely the Coulomb gauge and the light-front gauge. When the Coulomb gauge is considered as the natural one, one gets the Chen \emph{et al.} decomposition \cite{Chen:2008ag,Chen:2009mr} which we call the Coulomb GIE. When the light-front gauge is considered as the natural one, one gets the light-front GIE \cite{Hatta:2011ku,Lorce:2012ce} which reduces to the Bashinsky-Jaffe decomposition in the $A^+=0$ gauge \cite{Bashinsky:1998if,Wakamatsu:2010cb}. From the point of view of physically inequivalent GIEs, it is therefore not surprising that the canonical quantities obtained in the Coulomb GIE differ from those obtained in the light-front GIE \cite{Chen:2008ag,Wong:2010rs,Zhang:2011rn,Ji:2012sj}, see figure \ref{fig1} for an illustration. For example, Chen \emph{et al.} stress that the gluon spin appearing in their decomposition, \emph{i.e.} the Coulomb GIE, is not the $\Delta g$ measured in deep-inelastic scattering experiment \cite{Chen:2008ag,Chen:2011gn}. The latter corresponds in fact to the gluon spin appearing in the light-front GIE. 

Ji stressed that by adding gauge links one can transform gauge-variant quantities into gauge-invariant ones, but the price to pay is that the quantities become \emph{path dependent} and usually have a \emph{non-local} expression. Path dependence, which is just another way of saying that the GIEs are Stueckelberg variant \cite{Lorce:2012ce}, naturally raises the problem of uniqueness. Non-local expressions cannot easily be implemented in lattice calculations and have complicated transformation laws. Only in specific gauges they reduce to local expressions and have a simple interpretation. In other gauges, the interpretation becomes obscure. On the contrary, using the Chen \emph{et al.} approach one obtains operators which are both local and gauge invariant, leading to a clearer partonic interpretation. A typical example is the quantity called $\Delta g$ which is measured in deep-inelastic scattering experiments, and is usually associated with a non-local gauge-invariant operator \cite{Collins:1981uw}. This operator becomes local only in the light-front gauge, where it is naturally interpreted as the gluon spin contribution \cite{Manohar:1990kr,Manohar:1990jx,Balitsky:1991te,Jaffe:1995an}. In the light-front GIE, this $\Delta g$ is rewritten as a gauge-invariant local operator, see Eq. \eqref{Gspin}, clarifying its interpretation as the gluon spin in any gauge \cite{Wakamatsu:2010cb,Hatta:2011zs,Zhang:2011rn,Lorce:2012ce}. In this approach, the non-locality appears \emph{only when} one insists on writing the gauge-covariant potential $A^\text{phys}_\mu$ in terms of the gauge-variant potential $A_\mu$ and the field strength $F_{\mu\nu}$ \cite{Lorce:2012ce}. It might therefore be possible to compute $\Delta g$ on a lattice. The main difficulty consists in finding a way to implement $A^\text{phys}_\mu$, but this can be as problematic as simply fixing a gauge on the lattice. As a final remark, the complicated transformation law of the non-local operators can be understood as owing to the fact that the physical part of the gauge potential transforms as a Lorentz four-vector only up to a gauge transformation, as discussed in section \ref{LorentzSection}.

\subsection{Pragmatic point of view}

We have seen that there exist many acceptable decompositions from the point of view of gauge invariance, and accordingly many acceptable pictures. This makes some people feel uncomfortable. The reason is that if the physical ``reality'' is unique, there can be only one truly physical picture. The actual problem is therefore a problem of \emph{physical interpretation}. Which picture is the most ``physical'' is usually a matter of taste, a source of neverending debates, and is probably an \emph{ill-defined} problem. For these reasons, it is preferable to adopt a more pragmatic point of view. 

Many different physical pictures may coexist, as long as they are coherent and have a clear connection with physical observables. One has just to specify in which picture one is working. Depending on the situation, one picture may appear simpler and therefore more natural than the other ones. It does not mean that it has to be the ``physical'' one, but just the most \emph{convenient} one. For example, atomic systems are non-relativistic and are most conveniently described in the instant form of dynamics and the Coulomb gauge. For such systems, it is therefore more natural to work with the Coulomb GIE. On the contrary, the proton is a relativistic system and its internal structure is essentially probed in high-energy experiments involving large momentum transfer, where a parton model picture is very convenient \cite{Feynman:1969ej}. For this reason, it appears more natural to describe the proton in the framework of light-front dynamics \cite{Lepage:1980fj,Brodsky:1997de}. The contact with the parton model picture can then be achieved in the light-front gauge \cite{Jaffe:1989jz,Bashinsky:1998if}. In this context, it is clearly more convenient to work with the light-front GIE.

In summary, there are only two relevant decompositions for the proton spin puzzle:
\begin{itemize}
\item the Ji decomposition 
\begin{equation}
\frac{1}{2}=J_q+J_g,\qquad J_q=\frac{\Delta q}{2}+L_q
\end{equation}
with Wakamatsu's improvement based on the light-front GIE
\begin{equation}
J_g=\Delta g+L_g.
\end{equation}
\item the light-front GIE
\begin{equation}
\frac{1}{2}=\frac{\Delta q}{2}+\ell_q+\Delta g+\ell_g,
\end{equation}
which reduces to the Bashinsky-Jaffe decomposition in the light-front gauge.
\end{itemize}
The original Ji decomposition is the most conservative one in the sense that it does not require the existence of any preferred direction and any natural gauge. It is also the closest one to our classical picture where the particles follow well-defined trajectories. The main disadvantages of this decomposition are the absence of a simple partonic interpretation and the absence of a decomposition of the gluon angular momentum $J_g$ into spin and OAM contributions.  The only known way to solve the latter problem while preserving gauge invariance is to specify a preferred direction. High-energy experiments naturally break rotational symmetry by providing us with such a direction\footnote{By analogy, a Stern-Gerlach apparatus determines natural spin-up and spin-down states. The laws of physics are invariant by rotation but the experimental setup is not.}. This automatically makes the light-front gauge special and allows one to use a partonic picture. One can then define the projection of the gluon spin in that direction in a gauge-invariant way. Indeed, the $\Delta g$ extracted from DIS is associated with the $\mathsf M^{+12}_{g,\text{spin}}$ operator in the light-front GIE. Even though this gluon spin contribution appears naturally in the Bashinsky-Jaffe decomposition, Wakamatsu has shown that this piece of the puzzle can be used in a consistent way in the Ji decomposition \cite{Wakamatsu:2010cb}. In other words, the quark spin and gluon spin contributions are common to both the Bashinsky-Jaffe decomposition and the improved Ji decomposition.

In summary, there is no universal definition of the gluon spin. Part of the reason is that ``spin'' is too vague a word. Does it mean canonical polarization, helicity or light-front helicity? Each of these can be qualified as physical and there is fundamentally no reason to prefer one or another from theoretical arguments. The important question is which one can be accessed in experiments. The quantity $\Delta g$ measured in DIS has the physical interpretation of net gluon light-front helicity. The light-front GIE language simply makes this interpretation clear because it is well suited for describing light-front helicity.

\subsection{How to access the OAM?}

While the quark spin contribution is quite well determined and significant progress have been made in the determination of the gluon spin contribution \cite{deFlorian:2008mr}, very little is known about the OAM on the experimental side. It is however an essential piece  to solve the so-called ``spin puzzle'', see \emph{e.g.} \cite{Filippone:2001ux,Kuhn:2008sy}.  

Ji has shown that the quark and gluon kinetic angular momentum can be expressed in terms of twist-2 Generalized Parton Distributions (GPDs) \cite{Ji:1996ek} 
\begin{equation}\label{JiJz}
J^{q,g}=\frac{1}{2}\int\ud x\,x\left[H^{q,g}(x,0,0)+E^{q,g}(x,0,0)\right],
\end{equation}
which are used to describe some high-energy exclusive processes like \emph{e.g.} Deeply Virtual Compton Scattering and Deeply Virtual Meson Production. The quark and gluon kinetic OAM can then be obtained by subtracting, respectively, the quark and gluon spin contribution
\begin{equation}
L^q_{z}=J^q_z-\frac{\Delta q}{2},\qquad L^g_{z}=J^g_z-\Delta g.
\end{equation}
Note that these equations should not be considered as sum rules but rather as \emph{definitions} of the kinetic OAM.  While Ji's relation is valid in the target rest frame for all three components of the angular momentum, its derivation seems more natural for the transverse polarization in a leading-twist approach \cite{Burkardt:2002hr,Ji:2012sj}. The reason is that the OAM requires the correlation between the parton momentum and position. The GPDs encode the correlation between the longitudinal momentum and the transverse position \cite{Burkardt:2000za,Burkardt:2002hr,Burkardt:2005td}, and are therefore naturally related to the transverse polarization. Note that the transverse polarization raises some issue concerning the frame dependence of the decomposition into quark and gluon contributions \cite{Ji:2012vj,Hatta:2012jm,Leader:2012ar,Leader:2012md,Harindranath:2012wn}. The correlation involving the transverse momentum is encoded in higher twists. As first shown by Penttinen \emph{et al.}, the $z$-component of the quark kinetic OAM is related to a pure twist-3 GPD \cite{Penttinen:2000dg,Kiptily:2002nx}
\begin{equation}\label{PPSS}
L^q_z=-\int\ud x\,x\,G^q_2(x,0,0).
\end{equation}
The genuine spin sum rule in the quark sector is therefore given by
\begin{equation}
\int\ud x\left\{x\left[H^q(x,0,0)+E^q(x,0,0)+2G^q_2(x,0,0)\right]-\tilde H^q(x,0,0)\right\}=0,
\end{equation}
where we have used $\Delta q=\int\ud x\,\tilde H^q(x,0,0)$. Similar relations naturally hold in the gluon sector \cite{Hatta:2012cs,Ji:2012ba}.

The longitudinal component of the OAM involves the correlation between the parton transverse momentum and transverse position, which naturally leads to the concept of quantum phase-space or Wigner distribution, see \emph{e.g.} Refs. \cite{Wigner:1932eb,Ji:2003ak,Belitsky:2003nz,Lorce:2011kd}. The Wigner distribution is particularly intuitive as it is the closest quantum object to the classical concept of phase-space distribution. In particular, the quantum average of any operator $\widehat O(\widehat{\vec b}_\perp,\widehat{\vec k}_\perp,\widehat x)$ in a target state with polarization $\vec S$ can be expressed as a phase-space average
\begin{equation}
\langle\widehat O\rangle(\vec S)=\int\ud x\,\ud^2k_\perp\,\ud^2b_\perp\,O(\vec b_\perp,\vec k_\perp,x)\,\rho(\vec b_\perp,\vec k_\perp,x,\vec S),
\end{equation}
where $\rho(\vec b_\perp,\vec k_\perp,x,\vec S)$ is the Wigner distribution playing the role of a phase-space density, and $O(\vec b_\perp,\vec k_\perp,x)$ is the classical operator associated with the quantum operator $\widehat O(\widehat{\vec b}_\perp,\widehat{\vec k}_\perp,\widehat x)$. The Wigner distribution depends on the parton three-momentum $(\vec k_\perp,x=\tfrac{k^+}{P^+})$ and transverse position or impact parameter $\vec b_\perp$. Contrary to its non-relativistic version \cite{Wigner:1932eb,Ji:2003ak,Belitsky:2003nz}, the relativistic Wigner distribution has no dependence on the parton longitudinal position \cite{Lorce:2011kd}. The reason is that relativistic effects, like \emph{e.g.} Lorentz contraction, pair fluctuations and absence of relativistic concept of center of mass, spoil the semi-classical picture. These problems can however be avoided in the infinite-momentum frame where the target looks basically like a pancake \cite{Soper:1976jc,Burkardt:2000za,Burkardt:2002hr,Burkardt:2005td}. The orbital angular momentum then follows our classical intuition \cite{Lorce:2011kd,Lorce:2011ni}
\begin{equation}\label{OAMWigner}
\langle\widehat l_z\rangle(\vec e_z)=\int\ud x\,\ud^2k_\perp\,\ud^2b_\perp\,(\vec b_\perp\times\vec k_\perp)_z\,\rho(\vec b_\perp,\vec k_\perp,x,\vec e_z).
\end{equation}

To be gauge invariant, the operator definition of the Wigner distribution involves a gauge link. The consequence of this gauge link is that the Wigner distribution inherits a path dependence. Using a straight gauge link in Eq.~\eqref{OAMWigner} leads to the \emph{kinetic} OAM $L_z$ \cite{Ji:2012sj}. Note that for this to be true, the integration over $\vec k_\perp$ is crucial, which means that despite appearances the integrand $(\vec b_\perp\times\vec k_\perp)_z\,\rho(\vec b_\perp,\vec k_\perp,x,\vec e_z)$ does not represent a density of kinetic OAM \cite{Lorce:2012ce}. With the view of connecting the Wigner distributions to the Transverse-Momentum dependent parton  Distributions (TMDs) \cite{Meissner:2009ww,Lorce:2011dv} appearing in the description of high-energy semi-inclusive processes like Semi-Inclusive DIS and Drell-Yan \cite{Collins:1984kg,Ji:2004wu,Boer:2011fh}, it is more natural to consider instead a staple-like gauge link consisting of two longitudinal straight lines connected at $x^-=\pm\infty$ by a transverse straight line. In this case, Eq.~\eqref{OAMWigner} gives the \emph{canonical} OAM $\ell_z$ appearing in the light-front GIE \cite{Lorce:2011ni,Hatta:2011ku,Ji:2012sj}. Since $x^-$-independent gauge transformations leave the condition $A^+=0$ invariant, the light-front gauge does not completely fix the gauge in QCD \cite{Slavnov:1987yh,Antonov:1988yg,Mueller:1985wy}. This residual gauge freedom is the reason for the transverse gauge link at $x^-=\pm\infty$ which is crucial in the context of Single Spin Asymmetries \cite{Brodsky:2002cx,Brodsky:2002rv,Belitsky:2002sm}. Nevertheless, it has been shown to have no effects on $\Delta g$ and $\ell_z$ \cite{Bashinsky:1998if,Wakamatsu:2010cb,Hatta:2011zs,Hatta:2011ku,Hatta:2012cs,Zhang:2011rn,Ji:2012ba,Lorce:2012ce}. Some intutive picture has also been proposed \cite{Burkardt:2012sd}. Note that while the GPDs and the TMDs can be accessed in high-energy experiments, finding a process where the Generalized TMDs \cite{Meissner:2009ww}, which are connected via Fourier transform to Wigner distributions \cite{Lorce:2011kd}, naturally appear remains a big challenge. For completeness, let us also mention that a lot of efforts has been made to relate in a quantitative way the TMDs to the OAM \cite{Avakian:2008dz,Avakian:2010br,Avakian:2009jt,She:2009jq,Lorce:2011kn,Lu:2006kt,Bacchetta:2011gx}. So far, no model-independent relation has however been obtained. This is likely owing to the fact that the TMDs do not contain any information about the parton position.

\section{Conclusion}\label{ConclusionSection}

Chen \emph{et al.} proposed to decompose explicitly the gauge potential into pure-gauge and physical parts. We presented its geometrical interpretation and discussed its similarity with the famous Stueckelberg trick. It allows one to write gauge-invariant quantities satisfying the canonical commutation relations. Moreover, we argue that this approach is consistent with the Lorentz symmetry provided that one uses the appropriate Lorentz transformation law for the gauge potential. Thanks to the Chen \emph{et al.} approach, one can easily write down gauge-invariant extensions of the Jaffe-Manohar decomposition. The drawback is that there are as many gauge-invariant extensions as there are gauge conditions, raising the delicate question of deciding which is the ``physical'' one. The gauge invariance makes this problem ill-defined, and so experiments can just tell us which gauge condition is the most ``natural'' one in specific contexts. 

Since the proton structure is mainly probed in high-energy experiments  involving large momentum transfers, it seems quite natural to use the light-front gauge to make contact with the appealing parton model picture. There are therefore essentially two relevant spin decompositions: the Ji decomposition supplemented by the decomposition of the gluon angular momentum proposed by Wakamatsu, and the light-front gauge-invariant extension, which is a generalization of the Bashinsky-Jaffe decomposition to any gauge. They differ by a gauge-invariant term called the potential angular momentum representing the difference between the kinetic and the canonical angular momentum. The kinetic angular momentum corresponds to our classical intuition of angular momentum but does not have any simple partonic interpretation. On the contrary, the canonical orbital angular momentum has a simple partonic interpretation but requires the use of a natural gauge. The most intuitive approach to the orbital angular momentum in a relativistic quantum system is based on the concept of phase-space or Wigner distributions. Depending on the choice of the gauge link, one can obtain either the kinetic or the canonical orbital angular momentum. How to access these Wigner distributions remains a big challenge but would definitely improve drastically our understanding of the proton internal structure.

\acknowledgments

I benefitted from numerous discussions with the participants of the Workshop INT-12-49W on ``Orbital Angular Momentum in QCD'' that held from February 6th until February 17th 2012 at the Institute of Nuclear Theory of the University of Washington, Seattle. I would like to thank particularly Profs. E.~Leader, M.~Wakamatsu and F.~Wang for many helpful comments. This work was supported by the P2I (``Physique des deux Infinis'') network.

\end{document}